\documentclass[onecolumn,aps,prb,amsmath,amssymb,superscriptaddress]{revtex4}
\usepackage{epsfig}
\usepackage{graphicx}
\usepackage{dcolumn}
\usepackage{subfigure}
\usepackage{hyperref}
\usepackage{ulem}
\usepackage{bm}
\usepackage{color}
\usepackage{latexsym}
\hypersetup{bookmarks,
            colorlinks,
            filecolor=blue,
           urlcolor=blue,
           pdfpagemode=None}
\usepackage{latexsym}
\def \k{{\mathbf{k}}}
\def \q{{\mathbf{q}}}

\begin{document}
\title{Breakdown of Fermi liquid behavior at the $(\pi,\pi)=2$k$_F$ spin-density wave quantum-critical point: the case of electron-doped cuprates}

\pacs{PACS number}
\author{Dominic Bergeron}

\affiliation{D\'{e}partement de physique and RQMP, Universit\'{e} de Sherbrooke,
Sherbrooke, QC J1K 2R1, Canada}

\author{Debanjan Chowdhury}

\affiliation{Department of Physics, Harvard University, Cambridge MA 02138, U.S.A.}

\author{Matthias Punk}

\affiliation{Department of Physics, Harvard University, Cambridge MA 02138, U.S.A.}

\author{Subir Sachdev}

\affiliation{Department of Physics, Harvard University, Cambridge MA 02138, U.S.A.}

\author{A.-M.S. Tremblay}

\affiliation{D\'{e}partement de physique and RQMP, Universit\'{e} de Sherbrooke,
Sherbrooke, QC J1 2R1, Canada}
\affiliation{Canadian Institute for Advanced Research, Toronto, Ontario, Canada}

\begin{abstract}
Many correlated materials display a quantum critical point between a paramagnetic and a spin-density wave (SDW) state. The SDW wave vector connects points, so-called hot spots, on opposite sides of the Fermi surface. The Fermi velocities at these pairs of points are in general not parallel. Here we consider the case where pairs of hot spots coalesce, and the wave vector $(\pi,\pi)$ of the SDW connects hot spots with parallel Fermi velocities. Using the specific example of electron-doped cuprates, we first show that Kanamori screening and generic features of the Lindhard function make this case experimentally relevant. The temperature dependence of the correlation length, the spin susceptibility and the self-energy at the hot spots are found using the Two-Particle-Self-Consistent theory and specific numerical examples worked out for band and interaction parameters characteristic of the electron-doped cuprates. While the curvature of the Fermi surface at the hot spots leads to deviations from perfect nesting, the pseudo-nesting conditions lead to drastic modifications of the temperature dependence of these physical observables: Neglecting logarithmic corrections, the correlation length $\xi$ scales like $1/T$, namely $z=1$ instead of the naive $z=2$, the $(\pi,\pi)$ static spin susceptibility $\chi$ like $1/\sqrt T$, and the imaginary part of the self-energy at the hot spots like $T^{3/2}$. The correction $T_1^{-1}\sim T^{3/2}$ to the Korringa NMR relaxation rate is subdominant. We also consider this problem at zero temperature, or for frequencies larger than temperature, using a field-theoretical model of gapless collective bosonic modes (SDW fluctuations) interacting with fermions. The imaginary part of the retarded fermionic self-energy close to the hot spots scales as $-\omega^{3/2}\log\omega$. This is less singular than earlier predictions of the form $-\omega\log\omega$. The difference arises from the effects of umklapp terms that were not included in previous studies.
\end{abstract}
\date{\today}%
\maketitle
\section{Introduction}

Quantum phase transitions between a Fermi liquid and magnetic phases have been a subject of experimental and theoretical investigations for several decades~\cite{RoschRMP:2007,SachdevBook:2011,Mathur:1998,SachdevPS:2010}. The transition to a spin-density-wave (SDW) in particular is relevant to problems of current interest. In the cuprates, Daou \textit{et al.}~\cite{Daou:2009} argued that the Fermi surface change associated with this transition is a key to understanding anomalous normal state properties. Recent studies in the pnictides~\cite{Nandi:2010,Fernandes:2010,StewartRMP:2011}, heavy fermion materials~\cite{Knebel:2011} and organic superconductors~\cite{Sedeki:2009,Sedeki:2012,Bourbonnais:2008} focus on the relation between the SDW and superconductivity. In fact, strong experimental similarities between quantum critical behavior in the organics, pnictides and cuprates have been pointed out recently~\cite{Doiron:2009}.

The electron-doped cuprates~\cite{Armitage:2010} have provided an early example where quantum critical behavior has been inferred from the temperature dependence of resistivity at low temperature. It was measured to be linear ~\cite{Fournier:1998} from 35mK to 10K in Pr$_{2-x}$Ce$_x$CuO$_{4-\delta}$ (PCCO) at doping $x=0.17$. More recent transport~\cite{Dagan:2004} and thermopower measurements~\cite{Li:2007} also suggest the presence of a quantum critical point at a similar doping. The precise nature of the quantum critical point remains however unclear, as thoroughly discussed in Ref.~\onlinecite{Armitage:2010}. For example, it has also been suggested that the quantum critical point coincides with the onset of superconductivity in the overdoped regime~\cite{Jin:2011}.

\begin{figure}
\begin{center}
\includegraphics[width=0.9\columnwidth]{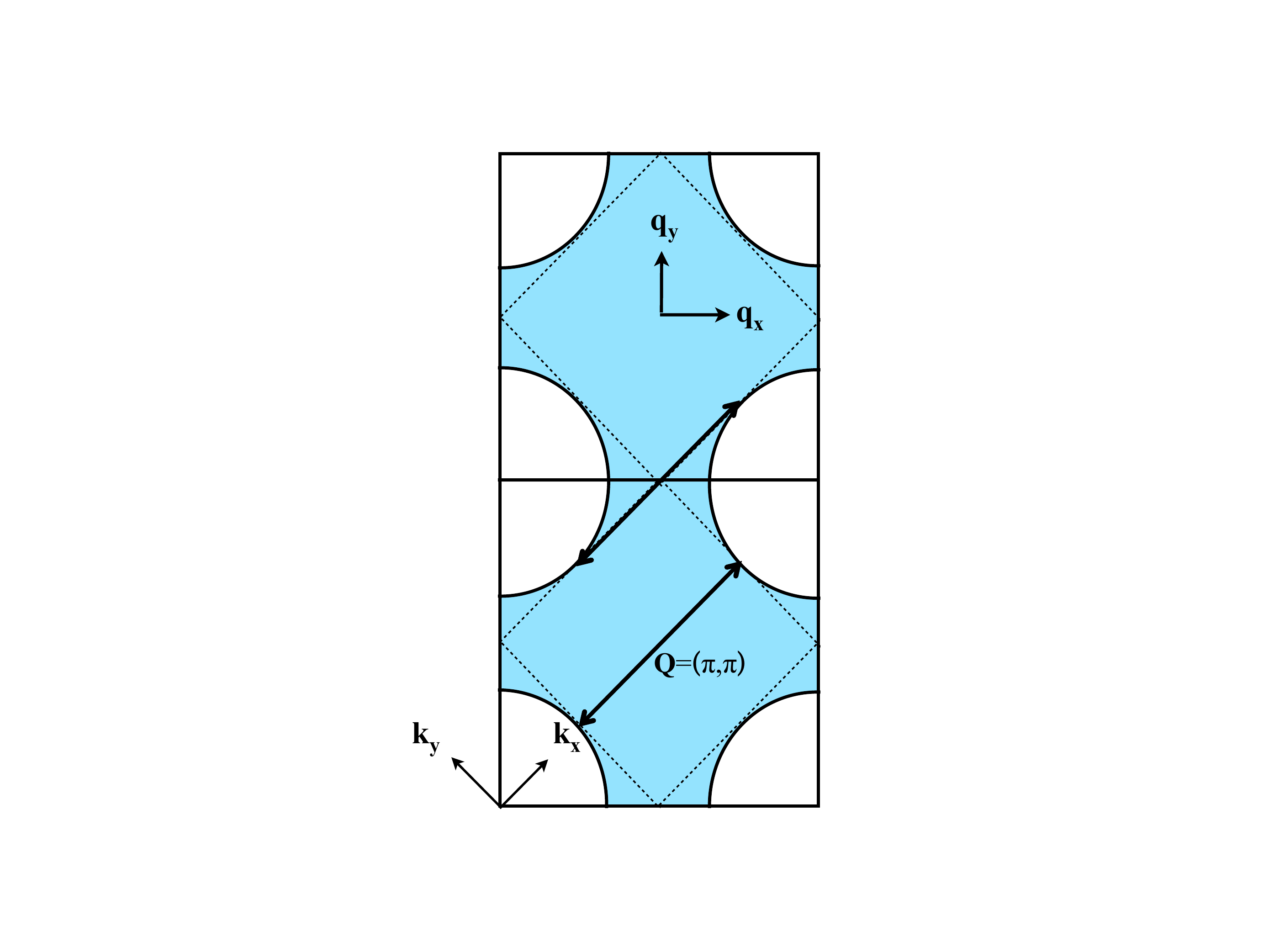}
\end{center}
\caption{The dashed lines indicate the magnetic Brillouin zone with ordering vector ${\bf Q}=(\pi,\pi)$. The arrows gives examples of pseudo-nesting conditions, namely of points such that $2{\bf k}_F$ is equal to the antiferromagnetic wave vector. The $(\pi,\pi)$ ordering wave-vector is defined with respect to the $(q_x,q_y)$ coordinate system. In the field theory approach introduced later, we work in the rotated ($k_x,k_y$) coordinate system.}
\label{bzf}
\end{figure}

Here we study quantum critical behavior associated with the transition between a SDW phase and a Fermi liquid when the wave vector $(\pi,\pi)$ of the SDW connects hot spots with parallel Fermi velocities. The two Fermi surfaces connected by $(\pi,\pi)$ in this case are tangent to each other, as shown in Fig.~\ref{bzf}. On a spherical Fermi surface, the Fermi wave vector would satisfy the condition $2{\bf k_F}=(\pi,\pi)$. We call this pseudo-nesting~\cite{Benard:1993}. We will explain why the QCP can be located at, or close to, the pseudo-nesting filling $n_c$. This occurs naturally in the one-band Hubbard model for the electron-doped cuprates and we will perform some of our calculations specifically for this case, although the frequency and temperature dependencies that we find are valid more generally. The methods that we describe below can be applied to electron-doped cuprates because these materials are described by a Hubbard model in an intermediate coupling regime where one can neglect effects induced by Mott physics~\cite{Senechal:2004,LTP:2006,PKAVC06,Weber:2010}.

The theory of Hertz~\cite{Hertz:1976,Beal:1975,RoschRMP:2007} and Millis\cite{Millis:1993} has formed the basis for much of the work on quantum critical phenomena. In this approach, fermions are integrated out and an effective bosonic theory for the collective modes is studied using standard renormalization group methods that can be taken to high order~\cite{Pankov:2004}. It has been pointed out by Abanov and Chubukov~\cite{Abanov:2003} that for a commensurate SDW at the upper critical dimension, namely $d=2$ for $z=2$, all the coefficients of the bosonic theory are singular so that one must treat simultaneously the bosonic collective modes and the fermions. Metlitski and Sachdev~\cite{MMSS10b} have reexamined this problem and obtained the non-Fermi liquid behavior at the hot spots, and shown that the bosonic SDW spectrum does not obey dynamic scaling with $z=2$ but instead that a super-power-law form is obtained. They have also thoroughly discussed the failure of the $1/N$ expansion at higher order, leading to a strong-coupling problem. More recently, it has been argued \cite{HHMS}
that non-Fermi liquid corrections are also important away from the hot spots.

An alternate approach is the self-consistent-renormalized theory of Moriya~\cite{Moriya:1985,Moriya:2006}. It is in the universality class of the spherical model and as such its critical behavior will not be exactly that expected for the O(3) model. Nevertheless, it can be accurate away from the critical point and provide leading order estimates for the exponents. The Two-Particle-Self-Consistent (TPSC) theory~\cite{Vilk:1997,TremblayMancini:2011} is a related approach that has critical behavior similar to that of Moriya, including logarithmic corrections~\cite{Roy:2008}. It has the advantage that although it is an approximate solution to the Hubbard model, it is quantitatively very close to benchmark Quantum Monte Carlo results~\cite{Vilk:1997}. It is non-perturbative, does not include any phenomenological parameters, has internal consistency checks, and satisfies a number of exact results.

Previous theoretical studies have mostly been done for the case where the quantum critical point occurs when the Fermi velocities at the hot spots that are connected by the SDW are not parallel, unlike the case of parallel velocities we consider here (see Fig.~\ref{bzf}). Such a case of parallel Fermi velocities is generic in one dimension, but at first sight appears as an accident in two dimensions, because upon translation by the $(\pi,\pi)$ SDW wave vector, the Fermi surfaces touch at only one point. If the surfaces were flat, we would recover the case of perfect nesting encountered in one dimension. The curvature here provides a cutoff, and so we refer to the situation with parallel Fermi velocities as ``pseudo-nesting''~\cite{Benard:1993}. Such Fermi surfaces have also been studied in three dimensions~\cite{Bazaliy:2004}, but in the presence of this pseudo-nesting the spin susceptibility is singular in two dimensions~\cite{Benard:1993} and the analysis must be redone. Altshuler, Ioffe and Millis~\cite{AIM95} first looked at the case where the instability is at $2k_F$, hence connects parallel segments of the Fermi surface, but the SDW wavelength is not commensurate with the lattice. They found that the transition is weakly first order with an intermediate scaling regime when the SDW wavelength is close to $(\pi,\pi)$. The scaling regime was obtained in a systematic expansion in a number proportional to the inverse number of fermion flavors. Krotkov and Chubukov~\cite{PKAVC06,Krotkov:2006a} found different results for the self-energy. Here we consider only the commensurate case. Some of our results differ from those of previous authors because they overlooked the significance of the umklapp process shown by the top double arrow in Fig.~\ref{bzf}.

We use two different approaches. We obtain finite temperature results appropriate for the SDW quantum critical point of electron-doped cuprates using TPSC. Then a field-theory for the spin-fermion model allows us to find the finite-frequency zero temperature results and some finite temperature results. The results of both approaches are consistent. We do not, however, consider the possibility of a first order transition \cite{AIM95}.

The rest of this paper is organized as follows. In Sec.~\ref{Sec:Model} we present the model along with general arguments suggesting why one should expect the antiferromagnetic quantum critical point to be located close to pseudo-nesting, namely at a filling where the $(\pi,\pi)$ wave vector connects parts of the Fermi surface that are tangent, or equivalently with parallel Fermi velocities. Sec.~\ref{Sec:Tfinite} contains the finite temperature results. They are obtained with TPSC, which is described in Sec.~\ref{Sub:TPSC}. Analytical results for the behavior at the QCP are illustrated with numerical examples appropriate for electron-doped cuprates in the subsections of Sec.~\ref{Sub:Analytical}. The critical behavior is obtained in Sec.~\ref{Sub:CriticalTPSC}. Zero temperature finite-frequency results are treated in Sec.~\ref{Sec:FieldTheory} with field theoretical methods. The Lagrangian appears in Sec.~\ref{Sub:Lagrangian} followed by sections on the polarization bubble (spin susceptibility)~\ref{Sub:PolarizationFT}, on the electron self-energy~\ref{Sub:SelfFT} and on the irrelevance of the quartic term~\ref{Sub:Scaling}. An appendix on vertex corrections~\ref{Sec:Vertex} appears after the summary in Sec.\ref{Sec:Summary}. Consistency between TPSC and field theory results are pointed out in the field theory Sec.~\ref{Sec:FieldTheory}.

\section{Model and QCP for electron-doped cuprates}\label{Sec:Model}

In this section, we introduce the model and give generic arguments why we expect the quantum critical point to often be located close to the filling where translation by the antiferromagnetic wave vector leads to Fermi surfaces that are tangent to each other, as illustrated in Fig.~\ref{bzf}.

While the frequency and temperature dependencies that we find do not depend on details of the model, specific numerical examples at finite temperature calculations will be performed on the two-dimensional $t-t^{\prime}-U$ Hubbard model on the square lattice at weak to intermediate coupling. The model is given by
\begin{equation}
H=-\sum_{\langle i,j\rangle,\sigma}t_{i,j}(c_{i,\sigma}^{\dagger}c_{j,\sigma
}+h.c.)+U\sum_{i}n_{i,\uparrow}n_{i,\downarrow}%
\end{equation}
where $t_{i,j}$ are the hopping integrals, $i,j$ are the site index, $\sigma$
is the spin label, $c_{i,\sigma}^{\dagger}$ and $c_{i,\sigma}$ are the
particle creation and annihilation operators. Doubly occupied sites cost
an energy $U$ and $n_{i,\sigma}=c_{i,\sigma}^{\dagger}c_{i,\sigma}$. Units
are such that $\hbar=1$, $k_{B}=1$ and lattice spacing is unity. The kinetic energy of a single-particle
excitation in momentum space is obtained from%
\begin{equation}
\varepsilon_{\mathbf{k}}=\left(  -\sum_{j}e^{i\mathbf{k\cdot}\left(
\mathbf{r}_{i}-\mathbf{r}_{j}\right)  }t_{i,j}\right)  -\mu^{\left(  1\right)
}%
\end{equation}
with the sum over $j$ running over all neighbors of any of the sites $i$. The chemical potential $\mu^{\left(  1\right)  }$ is chosen so that we have the required density.

\begin{figure}
  \centering
  \includegraphics[width=0.7\columnwidth]{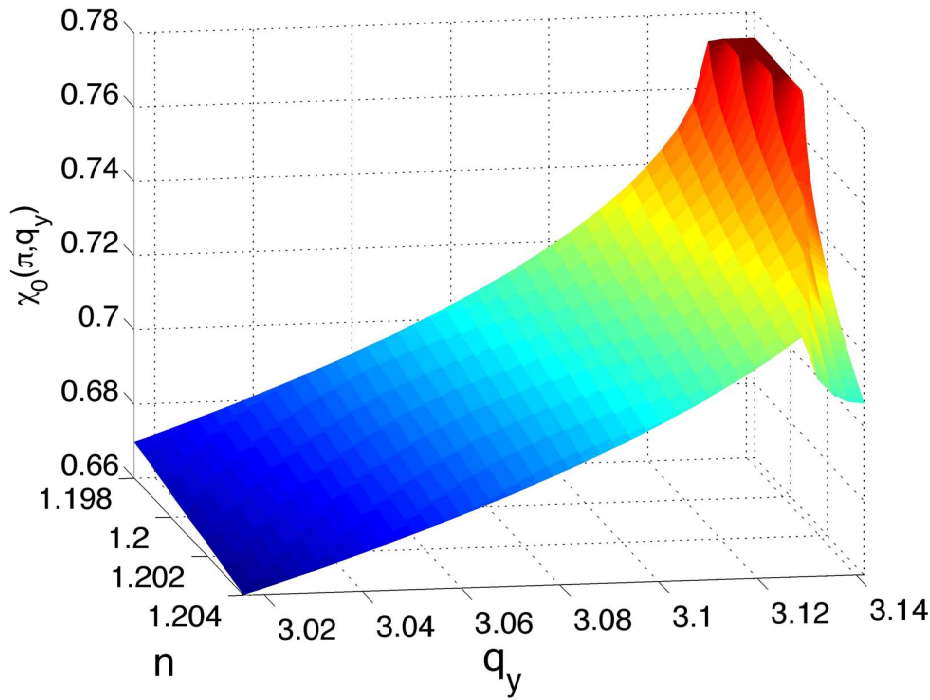}\\
  \caption{Lindhard function near the $2k_F$ point as a function of doping for $U=6$, $t^{\prime}=-0.175$ and $t^{\prime\prime}=0.05$, values that are appropriate for electron-doped cuprates. The rapid fall with filling larger than $1.201$, close to the critical filling, is apparent.}\label{Fig:SuscepAtQCP}
\end{figure}

One can explain on general grounds the filling where the QCP is likely to occur. Figure~\ref{Fig:SuscepAtQCP} displays the Lindhard function $\chi_0$ along the $q_y$ direction for different fillings. Its maximum is at $(\pi,\pi)$ so it is symmetric in $q_x$ and $q_y$. There are two remarkable features. First, below a certain doping $n_c$, the maximum value is almost independent of filling~\cite{Onufrieva:2004,Markiewicz:2003}, and second it falls rapidly as soon as the filling exceeds $n_c$. The filling $n_c$ corresponds to the point where the Fermi surfaces joined by $(\pi,\pi)$ touch instead of intersecting. A $0.3\%$ change in filling leads to almost $10\%$ drop in value of the susceptibility. If we consider a simple Stoner criterion for the transition, we would conclude that if $U$ takes the value $U_c=2/\chi_0^{max}~\sim 2.6$, then the QCP would be close to this filling $n_c$. This does not require fine tuning because the value of $U$ that should enter the Stoner criterion is the value renormalized by Kanamori-Br\"{u}ckner screening~\cite{Kanamori:1963,Brueckner:1960}. This renormalized value becomes essentially $U$ independent when $U$ becomes of the order of the bandwidth because the two-body wave function creates a cusp to minimize double-occupancy~\cite{Kanamori:1963,Brueckner:1960} and the renormalized interaction cannot become larger. This maximum renormalized value in TPSC, $U_{sp}$, takes a value~\cite{Vilk:1997} near $U_c\sim 2.6$. In addition, in TPSC the value of $U_{sp}$ self-consistently adjusts itself to the value necessary to prevent a finite temperature phase transition on the SDW side of the QCP. Although, at sufficiently low temperature, details will start to matter and one needs to start to tune the value of $U$ to find the QCP precisely at $n_c$, there is an intermediate temperature scale that can be quite broad where fine tuning is unnecessary.

\section{Finite temperature results and TPSC}\label{Sec:Tfinite}

In this section, we use the non-perturbative Two-Particle Self-Consistent (TPSC) approach \cite{Vilk:1997,TremblayMancini:2011}.
This approach respects the
Pauli principle, the Mermin-Wagner theorem and conversation laws. It also
contains quantum fluctuations in crossed channels that lead to
Kanamori-Br\"{u}ckner screening. \cite{Vilk:1994} It is valid in the weak to
intermediate coupling regime $\left(  U\lesssim6t\right)  $ and not too deep
in the renormalized classical regime where a pseudogap is observed. It has been benchmarked on
Quantum Monte Carlo calculations on the Hubbard model. \cite{Vilk:1997,
Vilk:1994, Vilk:1995, Veilleux:1995, Moukouri:2000, Kyung:2003a}.

TPSC has been shown to be in the $N=\infty$ universality class of the $O(N)$ model \cite{Dare:1996}. It has the same critical
behavior as Moriya theory and hence has the same logarithmic corrections~\cite{Roy:2008}.
These logarithms have the same functional form as those of the renormalization
group asymptotically close to the quantum critical point, but in TPSC and in
Moriya theory the mode-mode coupling term does not flow, hence the corrections
may differ in the details from the renormalization
group\ {\cite{RoschRMP:2007}}. Quantum critical behavior of the susceptibility
and of the self-energy in the closely related spin-fermion model has been
discussed by Abanov et al. \cite{Abanov:2003}.

It has been argued from detailed comparisons of numerical calculations with experiment~\cite{Kyung:2003,Senechal:2004,Hankevych:2006,LTP:2006,Krotkov:2006a} that strong-coupling physics is not important for electron-doped cuprates, at least not too close to half-filling. Hence, TPSC is appropriate to study these compounds. It gives a satisfactory description of ARPES data~\cite{Kyung:2004}, and the temperature $T^*$, where the pseudogap seen in ARPES opens up experimentally, corresponds to that where the antiferromagnetic correlation length coincides with the thermal de Broglie wavelength~\cite{Motoyama:2007}, as predicted for two-dimensional precursors of three-dimensional long-range order~\cite{Vilk:1995}.

Hence, all the numerical results are presented in units where $t=1$, $k_B=1$, $\hbar=1$ (with $z$ component of spin defined by $n_\uparrow -n_\downarrow$) for values of the Hubbard model hopping parameters appropriate for electron-doped cuprates, namely second and third nearest-neighbor hopping $t^{\prime}=-0.175$ and $t^{\prime\prime}=0.05$~\cite{Kyung:2004}. Interaction strengths $U=6$ and $U=5.56$, again in the range appropriate for electron-doped cuprates,~\cite{Kyung:2004} will be considered. 

We first present the formalism and then give analytical and numerical results for the QCP.

\subsection{TPSC}\label{Sub:TPSC}

Given the Hubbard model parameters, TPSC has no adjustable parameter. Irreducible vertices are obtained self-consistently and in such a way that the Pauli principle and conservation laws are obeyed. The formal derivation is given in Refs.~\onlinecite{Allen:2001,TremblayMancini:2011}. Here, we simply present the equations that are solved.

In TPSC, the retarded spin $\chi_{sp}(\mathbf{q},\omega)$ and charge $\chi_{ch}(\mathbf{q},\omega)$ susceptibilities are written as%
\begin{align}
\chi_{sp}(\mathbf{q},\omega) & = \frac{\chi^{\left(  1\right)  }(\mathbf{q}%
,\omega)}{1-\frac{U_{sp}}{2}\chi^{\left(  1\right)  }(\mathbf{q},\omega
)},
\label{Chisp}
\\
\chi_{ch}(\mathbf{q},\omega) & = \frac{\chi^{\left(  1\right)  }(\mathbf{q}%
,\omega)}{1+\frac{U_{ch}}{2}\chi^{\left(  1\right)  }(\mathbf{q},\omega
)},\label{Chich}
\end{align}
where $\chi^{\left(  1\right)  }(\mathbf{q},\omega)$ is the non-interacting
retarded Lindhard function at wave vector $\mathbf{q}$ and angular frequency
$\omega$%
\begin{equation}
\chi^{\left(  1\right)  }(\mathbf{q},\omega)=-\frac{2}{N}\sum_{\mathbf{k}%
}\frac{f\left(  \varepsilon_{\mathbf{k}}\right)  -f\left(  \varepsilon
_{\mathbf{k+q}}\right)  }{\omega+i\eta+\varepsilon_{\mathbf{k}}-\varepsilon
_{\mathbf{k+q}}}.\label{Lindhard}%
\end{equation}
Here, $f\left(  \varepsilon_{\mathbf{k}}\right)  $ is the Fermi function
$\left(  e^{\varepsilon_{\mathbf{k}}/T}+1\right)  ^{-1}$, $T$ is the temperature and $N$ is the total number of sites. The
effective spin interaction $U_{sp}$ is evaluated without adjustable parameter
using the \textit{ansatz\ }\cite{Vilk:1994,Vilk:1997}%
\begin{equation}
U\langle n_{\uparrow}n_{\downarrow}\rangle=U_{sp}\langle n_{\uparrow}%
\rangle\langle n_{\downarrow}\rangle
\end{equation}
with the local-moment sum rule that follows from the fluctuation-dissipation
theorem%
\begin{equation}
n-2\langle n_{\uparrow}n_{\downarrow}\rangle=\int_{-\infty}^{\infty}%
\frac{d\omega}{2\pi}\int_{-\infty}^{\infty}\frac{d^{2}q}{\left(  2\pi\right)
^{2}}\frac{2}{1-e^{-\omega/T}}\chi^{\prime\prime}_{sp}%
(\mathbf{q},\omega)\label{sum_rule}%
\end{equation}
where $\chi^{ \prime\prime}_{sp}=\operatorname{Im}\chi^{ \prime\prime}_{sp}$ and $\langle n_{\uparrow}n_{\downarrow
}\rangle$ is the double occupancy. We dropped the site index using translational
invariance and we used the Pauli principle to write
\begin{equation}
S^{2}\equiv\langle(n_{\uparrow}-n_{\downarrow})^{2}\rangle=n-2\langle
n_{\uparrow}n_{\downarrow}\rangle.
\end{equation}
Similarly the irreducible vertex $U_{ch}$ entering $\chi_{ch}(q)$ is found using a sum-rule that is the analog of Eqn.(\ref{sum_rule}) for spin: 

\begin{equation}
n+2\langle n_{\uparrow}n_{\downarrow}\rangle-n^2=\int_{-\infty}^{\infty}%
\frac{d\omega}{2\pi}\int_{-\infty}^{\infty}\frac{d^{2}q}{\left(  2\pi\right)
^{2}}\frac{2}{1-e^{-\omega/T}}\chi^{\prime\prime}_{ch}%
(\mathbf{q},\omega)\label{charge_sum_rule}%
\end{equation}

The crossing symmetric self-energy is obtained from \
\begin{equation}
\Sigma_{\sigma}^{\left(  2\right)  }(k)=Un_{-\sigma}+\frac{U}{8}\frac{T}%
{N}\sum_{q}\left[  3U_{sp}\chi_{sp}(q)+U_{ch}\chi_{ch}(q)\right]  G_{\sigma
}^{\left(  1\right)  }(k+q).\label{Self}%
\end{equation}
The superscript $\left(  2\right)  $ reminds us that we are at the second
level of approximation. $G_{\sigma}^{\left(  1\right)  }$ is the same Green's
function as that used to compute the susceptibilities $\chi^{\left(  1\right)
}(q)$. Charge fluctuations $\chi_{ch}(q)$ are included in numerical calculations
but they are neglected in the analytical results because they are small. 

Since the self-energy is constant at the first
level of approximation, this means that $G_{\sigma}^{\left(  1\right)  }$ is
the non-interacting Green's function with the chemical potential that gives
the correct filling. This chemical potential $\mu^{\left(  1\right)  }$ is
slightly different from the one that we must use in $\left(  G^{\left(
2\right)  }\right)  ^{-1}=i\omega_{n}+\mu^{\left(  2\right)  }-\varepsilon
_{\mathbf{k}}-\Sigma^{\left(  2\right)  }$ to obtain the same density
\cite{Kyung:2001}.

Unless otherwise specified, all the numerical results below are obtained using
the Matsubara frequency version of equations (\ref{Chisp}) to (\ref{Self}) without any approximation, hence they are valid at arbitrary distance from the
quantum critical point.

\subsection{Analytical results with numerical examples for electron-doped cuprates}\label{Sub:Analytical}

We begin below with the Ornstein-Zernicke form of the spin susceptibility that is usually valid when the correlation length is large. The case where there is perfect nesting leads us naturally to the pseudo-nesting condition relevant for this paper. The situation, however, is not as simple as usual since the Ornstein-Zernicke form for the spin susceptibility is incorrect in our case, as we will explain. The self-energy is treated at the end of this section.

\subsubsection{Ornstein-Zernicke form for the susceptibility}\label{Sub:SuscepTPSC}

\paragraph{General case}

When the correlation length is large, one usually assumes that the denominator of the
spin susceptibility can be expanded around the wave vectors $\mathbf{Q}_{d}$, where the
maxima in $\chi^{\left(  1\right)  }$ occur in $d-$dimensions. One then obtains%

\begin{equation}
\chi_{sp}^{\prime\prime}(\mathbf{q},\omega)=\frac{2}{U_{sp}\xi_{0}^{2}}%
\frac{\omega/\Gamma_{0}}{\left(  \xi^{-2}+\mathbf{q}^{2}\right)  ^{2}+\left(
\omega/\Gamma_{0}\right)  ^{2}},\label{Chi_As}%
\end{equation}
where $\mathbf{q}$ is measured with respect to the wave vector $\mathbf{Q}%
_{d}$ where the spin suseptibility is maximum ($(\pi,\pi)$ in our case). Defining $U_{mf}=2/\chi
^{\left(  1\right)  }\left(  \mathbf{Q}_{d}\mathbf{,}0\right)  $ as the value
of the interaction at the mean-field SDW transition, the other quantities
in the previous expression are
\begin{align}
\xi^{2} &  \equiv\xi_{0}^{2}\left(  \frac{U_{sp}}{\delta U}\right)
,\label{xsi2}\\
\delta U &  \equiv U_{mf}-U_{sp},\label{dU}\\
\xi_{0}^{2} &  \equiv-\frac{1}{2\chi^{\left(  1\right)  }\left(
\mathbf{q},0\right)  }\left.  \frac{\partial^{2}\chi^{\left(  1\right)
}\left(  \mathbf{q,}0\right)  }{\partial q^{2}}\right\vert _{\mathbf{0}%
},\label{xsi02}\\
\frac{1}{\Gamma_{0}} &  \equiv\frac{1}{\xi_{0}^{2}\chi^{\left(  1\right)
}\left(  \mathbf{q,}0\right)  }\left.  \frac{\partial\chi^{\left(  1\right)
\prime\prime}\left(  \mathbf{q},\omega\right)  }{\partial\omega}\right\vert
_{\omega=0}.\label{Gamma_0}%
\end{align}
In the expression for the spin susceptibility, the denominators are expanded
around the
$\left(  \pi,\pi\right)  $ wave vector.

To obtain analytical results for the imaginary part of the self-energy $\Sigma^{\left(  2\right)
\prime\prime R}\left(  \mathbf{k}_{F},\omega;T=0\right)  \ $ in Eqn.(\ref{Self}) we use the spectral representation for the susceptibilities and for the Green's function, perform the sum of the internal Matsubara frequency and then the analytical continuation neglecting the charge fluctuations, to obtain%

\begin{equation}
\Sigma^{\prime\prime R}\left(  \mathbf{k}_{F},\omega\right)  =-\frac{3UU_{sp}%
}{8}\frac{1}{2v_{F}}\int\frac{d^{d-1}q_{\Vert}}{\left(  2\pi\right)  ^{d-1}%
}\int\frac{d\omega^{\prime}}{\pi}\left[  n\left(  \omega^{\prime}\right)
+f\left(  \omega+\omega^{\prime}\right)  \right]  \chi_{sp}^{\prime\prime
}\left(  \mathbf{q}_{\Vert},q_{\perp}\left(  \mathbf{k}_{F}+\mathbf{Q}%
_{d},\omega,\omega^{\prime}\right)  ;\omega^{\prime}\right)
\label{ImSigmaReel}%
\end{equation}%
where $q_{\perp}$, the component of $\mathbf{q}$ parallel to the Fermi
velocity $\mathbf{v}_{F}$, is obtained from the solution of the equation
\begin{equation}
\varepsilon_{\mathbf{k}_{F}+\mathbf{Q}_{d}+\mathbf{q}}=\omega+\omega^{\prime
}.\label{Condition_q_par}%
\end{equation}
For all Fermi wave vectors, where $\varepsilon_{\mathbf{k}_{F}+\mathbf{Q}_{d}%
}\simeq0$ the above equation reduces to
\[
\mathbf{v}_{F}^{\prime}\cdot\mathbf{q}\simeq\omega+\omega^{\prime}%
\]
where $\mathbf{v}_{F}^{\prime}$ is the Fermi velocity in the hot region, i.e.
where $\varepsilon_{\mathbf{k}_{F}+\mathbf{Q}_{d}}\simeq0.$

In the asymptotic form of the spin susceptibility Eqn.(\ref{Chi_As}), the wave vector appears only in the form $\mathbf{q}_{\Vert}^2,q_{\perp}^2$ so that keeping this general form
in the equation for the self-energy Eqn.(\ref{ImSigmaReel}), we obtain%

\begin{widetext}
\begin{equation}
\Sigma^{\prime\prime R}\left(  \mathbf{k}_{F},\omega\right)
=-\frac{3UU_{sp}}{8}\frac{1}{2v_{F}}\int\frac{d^{d-1}q_{\Vert}}{\left(
2\pi\right)  ^{d-1}}\int\frac{d\omega^{\prime}}{\pi}\left[  n\left(
\omega^{\prime}\right)  +f\left(  \omega+\omega^{\prime}\right)  \right]
\chi_{sp}^{\prime\prime
}\left(  \mathbf{q}_{\Vert},(\omega+\omega')/v_F ;\omega^{\prime},T\right).
\label{SelfWithChipp}
\end{equation}
\end{widetext}%

Normally, one expects $\xi_{0}$ to be a temperature independent constant of
the order of the lattice spacing and $\Gamma_{0}/\xi_{0}^{2}$ to be a constant
of the order of the Fermi energy. In the case of perfect nesting, or of pseudo-nesting, this is not
the case.

\paragraph{Perfect nesting}%

\begin{widetext}%
Although the case we are interested in does not correspond to perfect nesting, understanding that case first will facilitate our task later. There is perfect nesting when the equality $\varepsilon
_{\mathbf{k}}=-\varepsilon_{\mathbf{k+Q}_{d}}$ is satisfied for all wave vectors, with $\mathbf{Q}_{d}$ the
nesting wave vector. This case was treated by Virosztek and
Ruvalds\cite{Virosztek:1990}. The quantities $\xi_{0}^{2}$ and
$\Gamma_{0}$ that are usually assumed temperature independent, here become
temperature dependent. We show this below.

For perfect nesting, the Lindhard function becomes%
\begin{equation}
\chi^{\left(  1\right)  }(\mathbf{Q}_{d},\omega)=\frac{2}{N}\sum_{\mathbf{k}%
}\frac{1-2f\left(  \varepsilon_{\mathbf{k}}\right)  }{\omega+i\eta
+2\varepsilon_{\mathbf{k}}}\label{Lindhard_nesting}%
\end{equation}
so that changing to an energy integral we have%
\begin{align}
\chi^{\left(  1\right)  \prime\prime}(\mathbf{Q}_{d},\omega) &  =-\pi\int
dEN_{d}\left(  E\right)  \left(  1-2f\left(  E\right)  \right)  \delta\left(
\omega+2E\right)  \label{ImLindhardNesting}\\
&  =\pi{N}_{d}\bigg(\frac{\omega}{2}\bigg)\text{tanh}\left(  \frac{\omega}{4T}\right)
\end{align}
where $N_{d}\left(  E\right)$ is the density of states. The real part at zero frequency on the other hand is given by%
\begin{equation}
\chi^{\left(  1\right)  }(\mathbf{Q}_{d},0)=\int dEN_{d}\left(  E\right)
\frac{\left(  1-2f\left(  E\right)  \right)  }{2E}=2\int_{0}^{\Lambda}%
dEN_{d}\left(  E\right)  \frac{\tanh\left(  E/2T\right)  }{2E}.%
\end{equation}

In two dimensions, there is a well known logarithmic divergence of the density
of states ${N}_{d}(\frac{\omega}{2})$ at the van Hove singularity. Neglecting this logarithmic divergence that appears only for a special filling in the hole-doped case, we
take ${N}_{d}(\frac{\omega}{2})$ as a constant. In that case, integrating by part and replacing the upper bound by infinity in the convergent integral we are left with
\begin{equation}
\chi^{\left(  1\right)  }(\mathbf{Q}_{d},0)  
= N_d\left(  0\right)  \left(
\left.  \ln (x)\tanh\left( x\right)  \right\vert _{0}^{\frac{\Lambda}{2T}
}-\int_{0}^{\infty}\frac{\ln x}{\cosh^{2}\left(x\right)
}dx\right)  \\
\approx  N_d\left(  0\right)  \ln\left(  \frac{\Lambda}{2T}\right)+B
\label{xhi1Log}%
\end{equation}
where $B$ is a temperature independent constant. We also have
\begin{equation}
\chi^{\left(  1\right)  \prime\prime}(\mathbf{Q}_{d},\omega)\approx\pi{N}%
_{d}(0)\text{tanh}\left(  \frac{\omega}{4T}\right)  .
\end{equation}
These results suggest that the quantity $\Gamma_{0}$ defined by Eqn.(\ref{Gamma_0})
scales as
\begin{equation}
\Gamma_{0}\sim\xi_{0}^{2}T\ln(\Lambda/T).\label{gamma0avecxsi0}%
\end{equation}

Following Ref.~\onlinecite{Dare:1996}$,$ we move on to demonstrate analytically that
$\xi_{0}^{2}$ in Eqn.(\ref{xsi02}) scales as $\xi_{0}^{2}\sim\frac{1}{T^{2}}$.
The $1/T^{2}$ dependence, fundamentally comes from the second derivative of
$\chi^{\left(  1\right)  }(\mathbf{Q}_{d},0)\approx N\left(  0\right)
\ln\left(  \frac{\Lambda}{T}\right)$ in Eqn.(\ref{xhi1Log}). We shall now make this argument
more rigorous. Keeping for a while a general notation where $i$ is
some direction in the Brillouin zone, and $\mathbf{q}$ is measured with
respect to the center of the zone, one can write
\begin{equation}
{\frac{\partial^{2}\chi^{\left(  1\right)  }\left(  \mathbf{q,}0\right)
}{\partial q_{i}^{2}}}=-2\int_{BZ}{\frac{d^{d}k}{(2\pi)^{d}}}{\frac
{\partial^{2}C}{\partial\epsilon_{\mathbf{k+q}}^{2}}}\ \left(  \frac
{\partial\epsilon_{\mathbf{k+q}}}{\partial q_{i}}\right)  ^{2}-2\int
_{BZ}{\frac{d^{d}k}{(2\pi)^{d}}\frac{\partial C}{\partial\epsilon
_{\mathbf{k+q}}}}\frac{\partial^{2}\epsilon_{\mathbf{k+q}}}{\partial q_{i}%
^{2}},\label{a4}%
\end{equation}%
\end{widetext}%
where
\[
C(\epsilon_{\mathbf{k+q}},\epsilon_{\mathbf{k}})={\frac{f(\epsilon
_{\mathbf{k+q}})-f(\epsilon_{\mathbf{k}})}{\epsilon_{\mathbf{k+q}}%
-\epsilon_{\mathbf{k}}}.}%
\]
Measuring $\mathbf{q}$ with respect to $\mathbf{Q}_{d}$ we evaluate the above
second derivative at $\mathbf{q}=0.$ As before, for perfect nesting we have
\begin{align}
C(\epsilon_{\mathbf{k+Q}_{d}+\mathbf{q}},\epsilon_{\mathbf{k}})  &
=\frac{2f\left(  \epsilon_{\mathbf{k+\mathbf{Q}_{d}+q}}\right)  -1}%
{2\epsilon_{\mathbf{k+\mathbf{Q}_{d}+q}}}=-\frac{\tanh\left(  \epsilon
_{\mathbf{k+\mathbf{Q}_{d}+q}}/2T\right)  }{2\epsilon_{\mathbf{k+\mathbf{Q}%
_{d}+q}}}\\
& \equiv\frac{1}{T}F\left(  \frac{\epsilon_{\mathbf{k+\mathbf{Q}_{d}+q}}}%
{T}\right)  .
\end{align}
The last equation shows that $C$ scales as $T^{-1}$ times a function of
$\epsilon_{\mathbf{k+\mathbf{Q}_{d}+q}}/T.$ In the integrals, the derivatives
of the type $\frac{\partial\epsilon_{\mathbf{k+q}}}{\partial q_{i}}$ will not
introduce singular terms in temperature. Hence, replacing them by some average
value in the Brillouin zone, we can change the integration variable to energy
and the most singular terms in temperature will come from
\begin{align}
\left.  \frac{\partial^{2}\chi^{\left(  1\right)  }\left(  \mathbf{q,}%
0\right)  }{\partial q_{i}^{2}}\right\vert _{\mathbf{q=0}}  & \simeq-\int
{dEN}_{d}\left(  E\right)  {\frac{\partial^{2}C}{\partial E^{2}}}\ \left(
\mathbf{v}_{i}\right)  ^{2}\label{d2chidq2initial}\\
& =\frac{1}{T}\int{dEN}_{d}\left(  E\right)  \frac{\partial^{2}F\left(
\frac{E}{T}\right)  }{\partial E^{2}}\left(  \mathbf{v}_{i}\right)  ^{2}%
\end{align}
Neglecting the energy dependence of the density of states, we are left with%
\begin{equation}
\left.  \frac{\partial^{2}\chi^{\left(  1\right)  }\left(  \mathbf{q,}%
0\right)  }{\partial q_{i}^{2}}\right\vert _{\mathbf{q=0}}=\frac{1}{T^{2}}%
\int{dxN}_{d}\left(  0\right)  \frac{\partial^{2}F\left(  x\right)  }{\partial
x^{2}}\left(  \mathbf{v}_{i}\right)  ^{2}.\label{d2chidq2}%
\end{equation}
Using the definition of $\xi_{0}^{2}$, Eqn.(\ref{xsi02}), and the result for
$\Gamma_{0}$ Eqn.(\ref{gamma0avecxsi0}) above, we have that
\begin{equation}
\xi_{0}^{2}\sim\frac{1}{T^{2}}\;;\;\Gamma_{0}\sim\frac{\ln(\Lambda/T)}{T}%
.\label{xsi0andGamma0Perfect}%
\end{equation}

\paragraph{Pseudo-nesting}

\begin{figure}
  \centering
  \includegraphics[width=0.4\columnwidth]{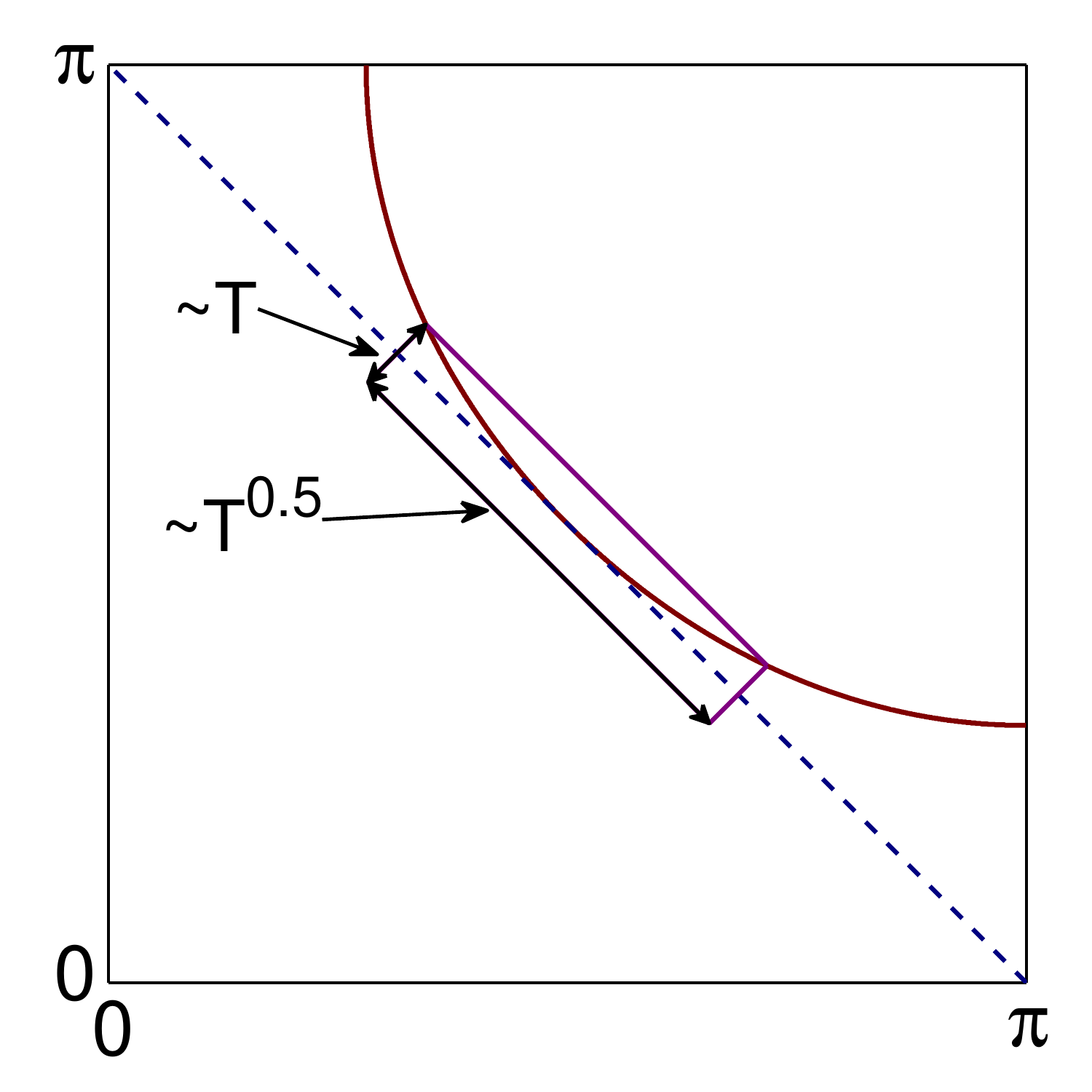}\\
  \caption{The Fermi surface at the critical filling $n_c$ touches the antiferromagnetic Brillouin zone. The important integration region is over a rectangle of thickness $T$ and width $\sqrt T$. The critical chemical potential $\mu_c$ where the antiferromagnetic zone boundary touches $(\pi/2,\pi/2)$ on the Fermi surface is the solution of $-2t''(\cos(\pi)+\cos(\pi))-\mu_c=0$. This corresponds to a filling $n_c=1.2007$ for $t'=-0.175$ and $t''=0.05$. For $U=6$, the critical filling that we find, $n_c=1.20096$, is slightly larger.
}\label{Fig:IntegrationDomain}
\end{figure}


In this subsection we show that, for the pseudo-nesting case, the main contribution to the Lindhard function at $\mathbf{Q}_d$ has the same form as in the perfect nesting case except for a temperature dependant prefactor. The previous calculation illustrates that the main contribution to the quantities of interest, $\xi_{0}^{2}$ and $\Gamma_{0},$ come from nested regions of the Fermi surface. In the pseudo-nesting case~\cite{Benard:1993} illustrated in Fig.~\ref{Fig:IntegrationDomain}, the Fermi surface displaced by $\mathbf{Q}_d$ just touches the original Fermi surface, with the Fermi velocities of the two surfaces that are parallel at the touching point. As in the perfect nesting case, the most important contribution to the integral for the Lindhard function around $\mathbf{Q}_d$ comes from the regions in $k$-space connected by $\mathbf{Q}_d$, with a width around  the Fermi surface that corresponds to an energy range $\varepsilon_{\mathbf{k}}\simeq T$. Now, imagine that we divide the integral over $\mathbf{k}$ near one of those points of the Fermi surface, for example $\mathbf{k}_F^{Q}=(\frac{\pi}{2},\frac{\pi}{2})$, into two components, $\mathbf{k}_{\perp}$ parallel to the Fermi velocity, i.e. perpendicular to the Fermi surface, and $\mathbf{k}_{||}$ parallel to the Fermi surface at $\mathbf{k}_F^{Q}$. For $\mathbf{k}$ near that region of the Fermi surface, we can write in two dimensions
\begin{equation}\label{eq:approx_eps_k}
\varepsilon_{\mathbf{k}_{||}+\mathbf{k}_{\perp}}\simeq\left(\frac{\partial \varepsilon_{\mathbf{k}_{||}+\mathbf{k}_{\perp}}}{\partial\mathbf{k}_{\perp}}\right)  \delta k_{\perp}+\frac{1}{2}\frac{\partial^{2}\varepsilon_{\mathbf{k}_{||}+\mathbf{k}_{\perp}}}{\partial\mathbf{k}_{||}^{2}}\delta k_{_{||}}^{2}\simeq v_{F}\delta k_{\perp}+\frac{1}{2}\kappa\delta
k_{_{||}}^{2}\,,
\end{equation}
where we have measured wave vectors with respect to $\mathbf{k}_F^{Q}$ and used the fact that $\nabla_{\mathbf{k}_{||}}\varepsilon_{\mathbf{k}_{||}+\mathbf{k}_{\perp}}=0.$ The quantity $\kappa$ measures the curvature of the Fermi surface. From that approximation we have $\varepsilon_{\mathbf{k}_{||}+\mathbf{k}_{\perp}+\mathbf{Q}_{d}}\simeq-v_{F}\delta k_{\perp}+\frac{1}{2}\kappa\delta k_{||}^{2}$ and thus
\begin{equation}\label{eq:diff_eps_k}
\varepsilon_{\mathbf{k}_{||}+\mathbf{k}_{\perp}}-\varepsilon_{\mathbf{k}_{||}+\mathbf{k}_{\perp}+\mathbf{Q}_{d}}\simeq 2v_F \delta k_{\perp}\,.
\end{equation}
Since the terms in $\delta k_{||}^{2}$ cancel out in that expression, this approximation is valid if the next term in the series of $\varepsilon_{\mathbf{k}_{||}+\mathbf{k}_{\perp}}$ is negligible compared with the first one, namely if
\begin{equation}
\left\vert\frac{1}{2}\frac{\partial^2 \varepsilon_{\mathbf{k}_{||}+\mathbf{k}_{\perp}}}{\partial k_{\perp}^2} \delta k_{\perp}^2\right\vert \ll \left\vert v_F \delta k_{\perp}\right\vert
\end{equation}
and since $v_F \delta k_{\perp}\lesssim T$, we have the following upper bound for the temperature,
\begin{equation}
T \ll \left\vert \frac{v_F^2}{\frac{1}{2}\frac{\partial^2 \varepsilon_{\mathbf{k}_{||}+\mathbf{k}_{\perp}}}{\partial k_{\perp}^2} }\right\vert\,.
\end{equation}
Now, for the Fermi function, over a region around $\mathbf{k}_F^{Q}$, we have
\begin{equation}
\begin{split}
f\left(  \varepsilon_{\mathbf{k}}\right)    & \simeq f\left(  v_{F}\delta
k_{\perp}\right)  +\left.  \frac{\partial f\left(  \varepsilon\right)  }%
{\partial\varepsilon}\right\vert _{\varepsilon=v_{F}\delta k_{\perp}}\left(
\frac{1}{2}\kappa\delta k_{_{||}}^{2}\right) + ... \\
f\left(  \varepsilon_{\mathbf{k+Q}_{d}}\right)    & \simeq f\left(
-v_{F}\delta k_{\perp}\right)  +\left.  \frac{\partial f\left(  \varepsilon
\right)  }{\partial\varepsilon}\right\vert _{\varepsilon=-v_{F}\delta k_{\perp}%
}\left(  \frac{1}{2}\kappa\delta k_{_{||}}^{2}\right) + ...
\end{split}
\end{equation}
Since the derivative of the Fermi function is even in $\varepsilon$,
\begin{equation}\label{eq:diff_Fermi_eps}
f\left(  \varepsilon_{\mathbf{k}}\right)-f\left(  \varepsilon_{\mathbf{k+Q}_{d}}\right)\simeq 2f(v_{F}\delta k_{\perp})-1\,.
\end{equation}
The region where this is valid is given by the condition
\begin{equation}
\left\vert \left. \frac{1}{2} \frac{\partial^2 f\left(  \varepsilon\right)  }%
{\partial\varepsilon^2}\right\vert _{\varepsilon=v_{F}\delta k_{\perp}}\left(
\frac{1}{2}\kappa\delta k_{_{||}}^{2}\right)^2  \right\vert \ll f\left(
v_{F}\delta k_{\perp}\right)\,,
\end{equation}
which is satisfied if $\kappa\delta k_{_{||}}^{2}\ll T$. Therefore, from \eqref{eq:diff_eps_k} and \eqref{eq:diff_Fermi_eps}, the Lindhard function takes the form
\begin{equation}
\chi^{\left(  1\right)  }(\mathbf{Q}_{d},\omega)=\frac{2}{N}\sum_{k_{||}\in D_{2}}\sum_{k_{\perp}\in D_{1}}\frac{1-2f\left(  v_{F}\delta k_{\perp}\right)  }{\omega+i\eta+2v_{F}\delta k_{\perp}}\quad+\quad \text{less singular} \label{Lindhard_pseudo-nesting}
\end{equation}
where $D_{1}$ is a domain such that $v_{F}\delta k_{\perp}\lesssim T$ while $D_{2}$
is the domain such that $\kappa\delta k_{_{||}}^{2}\lesssim$
$T.$ The integration over $D_{2}$ thus gives a factor proportional to $\sqrt{T}$. The integral over $\delta k_{\perp}$ can be transformed into an
integral over energy in the same way as the perfect nesting case, with a
constant density of states determined by the Fermi velocity. The domain delimited by $D_{1}$ and $D_{2}$ is depicted in Fig.~\ref{Fig:IntegrationDomain}.

Overall then, the final result will be that
\begin{equation}
\chi^{\left(  1\right)  \prime\prime}(\mathbf{Q}_{d},\omega)\sim T^{1/2}%
\tanh\left(  \frac{\omega}{T}\right)\label{Chi''(1)0}
\end{equation}
where the $T^{1/2}$ prefactor comes from the $k_{_{||}}$ integration.
A similar reasoning leads to%
\begin{equation}
\chi^{\left(  1\right)  }(\mathbf{Q}_{d},0)\approx T^{1/2}\ln\left(
\frac{\Lambda}{T}\right)+A\label{Chi(1)0}
\end{equation}
which means that the regular temperature independent term represented here by $A$ dominates at low temperature. 

Repeating the same analogous arguments for ${\frac{\partial^{2}\chi^{\left(
1\right)  }\left(  \mathbf{q,}0\right)  }{\partial q_{i}^{2}},}$ we find that
\begin{equation}
\xi_{0}^{2}\sim\frac{T^{1/2}}{T^{2}}\sim\frac{1}{T^{3/2}}
\label{Xsi_0}
\end{equation}
which implies from the definition of $\Gamma_0$ Eqn.(\ref{Gamma_0}) and Eqs.(\ref{Chi''(1)0}-\ref{Chi(1)0}) that
\begin{equation}
\Gamma_{0}\sim\xi_{0}^{2}T^{1/2}\sim\frac{1}{T},
\end{equation}%
the same result for $\Gamma_{0}$, within logarithmic corrections, as if we had perfect nesting. In three
dimensions, the correction compared to perfect nesting is determined by the area spanned by $\delta k_{||}$, proportional to $\delta k_{||}^2\sim T$, hence we would have had $\xi_{3d,0}^{2}\sim\frac{T}{T^{2}}$ and again $\Gamma_{3d,0}\sim\frac{1}{T}.$

\begin{figure}[ht]
  \centering
   \begin{minipage}[c]{.46\linewidth}
     \includegraphics[width=0.9\columnwidth]{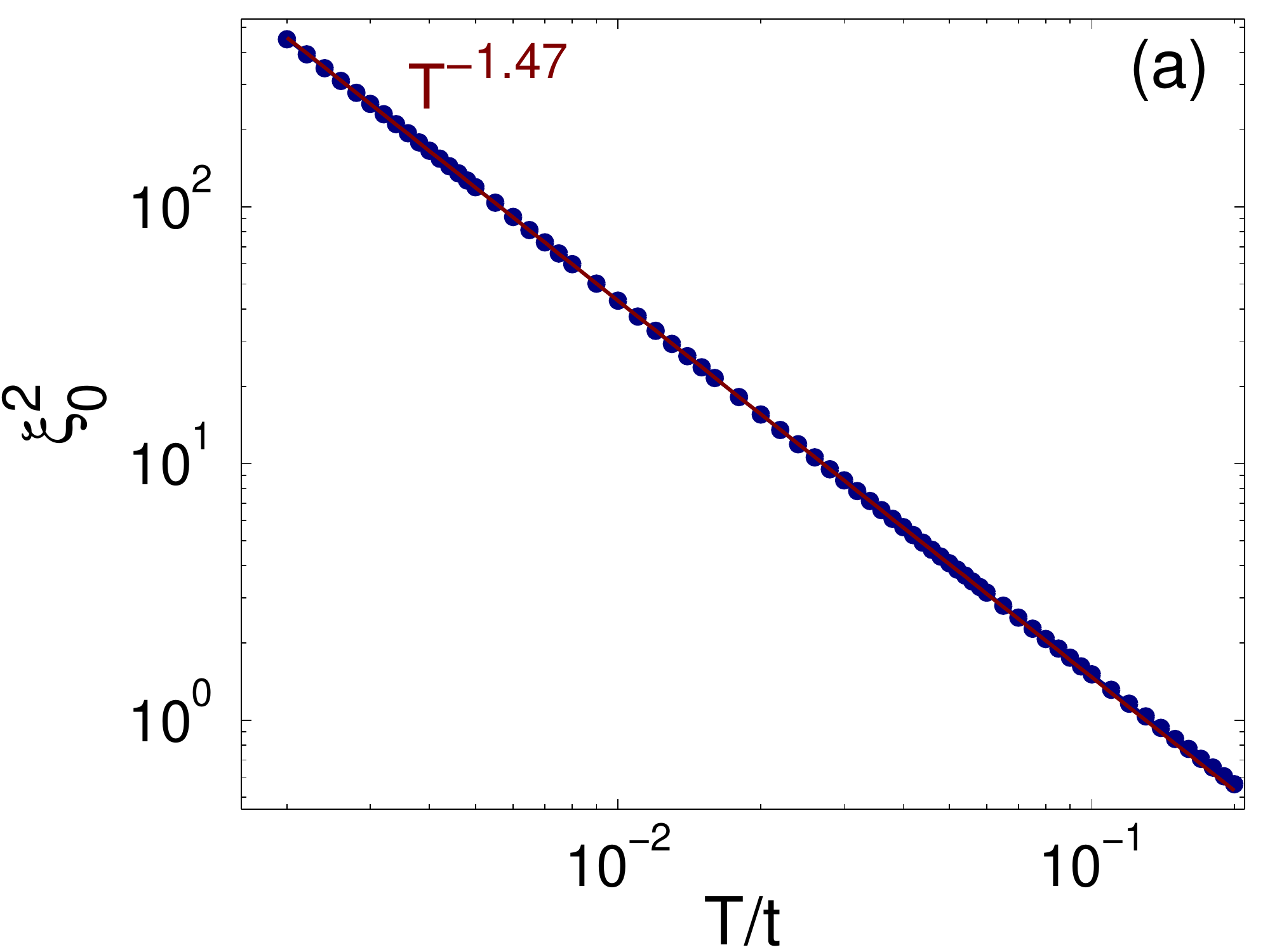}
   \end{minipage} \hfill
  \begin{minipage}[c]{.46\linewidth}
    \includegraphics[width=0.9\columnwidth]{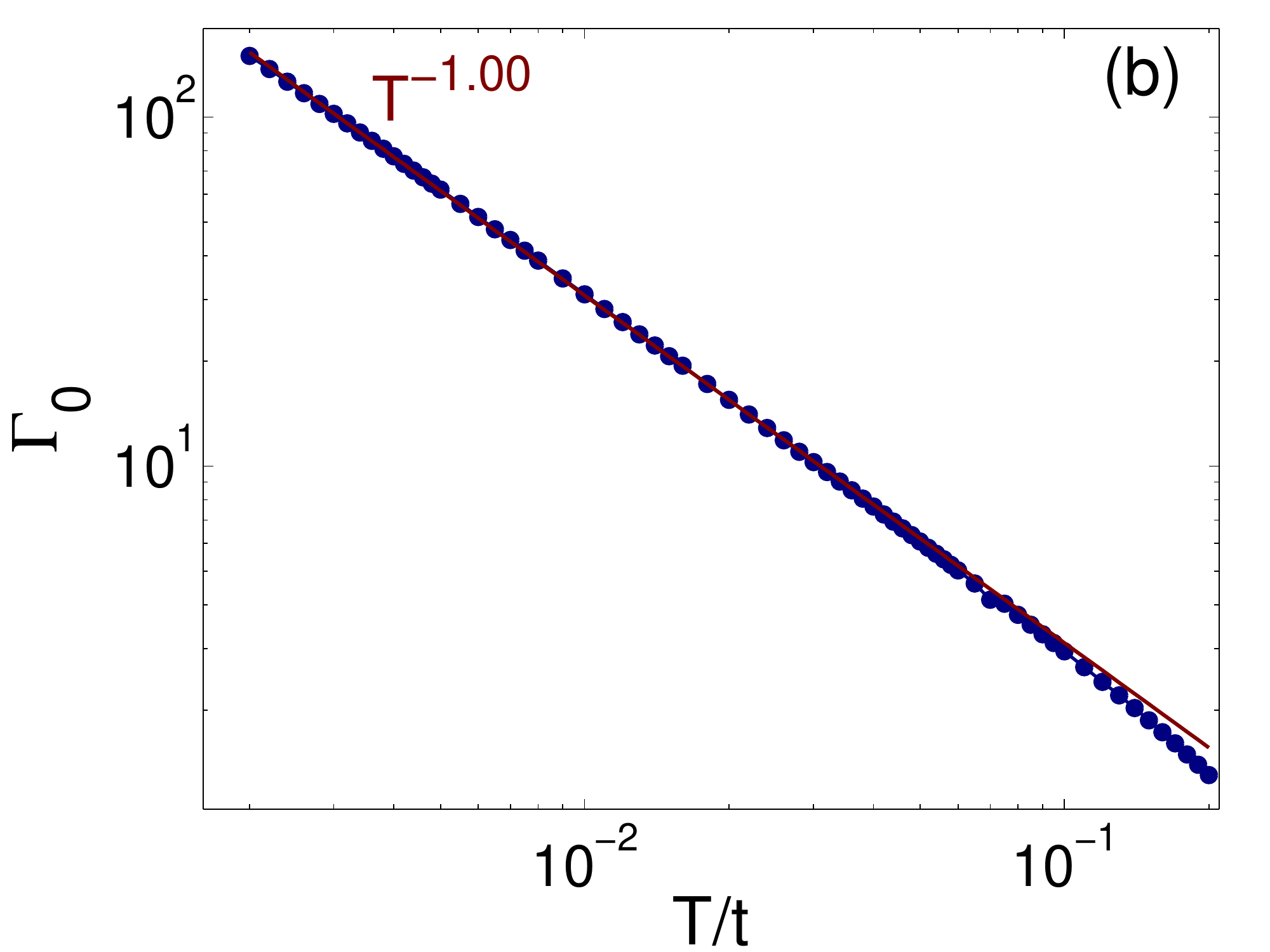}
   \end{minipage}

  \caption{$\xi_0$ and $\Gamma_0$ evaluated at $n_c=1.2007$ obtained from the condition that the Fermi surface is tangent to the antiferromagnetic zone boundary, as explained in the caption of Fig.~\ref{Fig:IntegrationDomain}.}
  \label{Fig:Xsi0_and_Gamma_0}
\end{figure}

Results of numerical calculations shown in Fig.~\ref{Fig:Xsi0_and_Gamma_0} confirm the power law temperature dependencies found above. At the actual filling $n_c$ where the Fermi surface is tangent to the antiferromagnetic zone boundary, the power laws extend to low temperature. 

\subsubsection{Higher order and scaling}

Given that $\xi_{0}$ and $\Gamma_{0}$ are now temperature dependent, we should
check whether the small $\mathbf{q}$ and small $\omega$ expansion of the
denominator that lead to Eqn.(\ref{Chi_As}) is still valid.
Normally, the expansion is of the form
\begin{equation}
\chi_{sp}^{\prime\prime}(\mathbf{q},\omega)=\frac{2\xi^{2}}{U_{sp}\xi_{0}^{2}%
}\operatorname{Im}\frac{1}{1+\mathbf{q}^{2}\xi^{2}+a\mathbf{q}^{4}\xi^{2}-i\omega\xi^{2}/\Gamma_{0}},
\label{Suscep_simple}
\end{equation}
with $a$ a constant. Since the function falls on a scale $\mathbf{q}^{2}\sim \xi^{-2}$ the higher order term $a\mathbf{q}^{4}\xi^{2}\sim \xi^{-2}$ can be neglected. However, our case is different. The coefficients of the expansion in powers of $\mathbf{q}$ are singular at $T=0$. For example we have $\left.  \frac{\partial^{4}\chi^{\left(  1\right)  }\left(  \mathbf{q,}0\right)  }{\partial q_{i}^{4}}\right\vert _{\mathbf{q=0}} \sim\frac{T^{1/2}
}{T^{4}}$ where the $1/T^4$ comes from counting the powers of $T$ associated with derivatives in Eqs.(\ref{d2chidq2initial}) to (\ref{d2chidq2}) and the $T^{1/2}$ from the restriction to the $k_{||}$ integral as usual. Knowing the scaling of $\xi_{0}^{2}$, we can rewrite 
$\left.  \frac{\partial^{4}\chi^{\left(  1\right)  }\left(  \mathbf{q,}0\right)  }{\partial q_{i}^{4}}\right\vert _{\mathbf{q=0}}\sim\xi_{0}^{2}\frac{1}{T^{2}}$ so that
we are left with%
\begin{equation}
\chi_{sp}^{\prime\prime}(\mathbf{q},\omega) =\frac{2\xi^{2}}{U_{sp}\xi
_{0}^{2}}\operatorname{Im}\frac{1}{1+\mathbf{q}^{2}\xi^{2}+\frac{a'}{T^{2}%
}\mathbf{q}^{4}\xi^{2}-i\omega\xi^{2}/\Gamma_{0}}\\
\end{equation}
where $a'$ is a constant.  The susceptibility will preserve a scaling form as a function of $q/T$ and $\omega/T$ if the scaling exponent is $z=1$. Indeed, in that case $\xi\sim1/T$ and since $\Gamma_{0}\sim1/T,$ $\xi_{0}^{2}\sim1/T^{3/2}$, the susceptibility becomes
\begin{equation}
\chi_{sp}^{\prime\prime}(\mathbf{q},\omega) \sim\frac{1}{T^{1/2}}\operatorname{Im}\frac{1}{1+\frac{b'\mathbf{q}^{2}}%
{T^{2}}+\frac{c'\mathbf{q}^{4}}{T^{4}}-i\frac{d'\omega}{T}}%
\end{equation}
with $b', c', d'$ constants. Each higher power of $\mathbf{q}^2$ has an additional power of $1/T^2$ coming from the additional derivatives of the non-interacting susceptibility and the scaling form is preserved to all orders. For frequency, there are also higher order terms, $\left(
\omega/T\right)^{3}$ etc. Hence we have the general scaling form
\begin{equation}
\chi_{sp}^{\prime\prime}(\mathbf{q},\omega) = \frac{1}{\sqrt{T}}g\left(\frac{q}{T},\frac{\omega}{T}\right).
\label{ChiSpScaling}
\end{equation}
We check in the following section that this is consistent with the TPSC self-consistency condition.

\subsection{Quantum Critical behavior}\label{Sub:CriticalTPSC}
\subsubsection{Correlation length, spin susceptibility and NMR relaxation rate}
The quantum critical behavior has been thoroughly studied in Ref.\onlinecite{Roy:2008} for the case where the Ornstein-Zernicke form is valid. This analysis does not apply here because of the more general form of the spin susceptibility obtained in the previous section. Nevertheless, it is interesting to note that with the Ornstein-Zernicke form, one obtains  
\begin{equation}
\xi^{-2}\sim\frac{T}{\Gamma_{0}}\sim T^{2}%
\label{Xsi_QCP}
\end{equation}
for both perfect and pseudo nesting. Hence, simply taking into account the temperature dependence of $\Gamma_0$, we recover $z=1$ scaling, namely%
\begin{equation}
\xi\sim\frac{1}{T}.%
\end{equation}
Note that with a temperature independent value for $\Gamma_0$ we recover the more usual result~\cite{Roy:2008,Moriya:1990} $z=2$.

The physics of the result for the correlation length is however quite different from the calculation with the Ornstein-Zernicke form. Indeed, in the latter case, it is the self-consistency relation Eqn.(\ref{sum_rule}) that determines the temperature dependence of the correlation length. In the present case, we found that temperature dependence in the previous section without invoking the self-consistency relation. We will show in Sec.(\ref{Sub:Scaling}) that indeed in our case, the self-consistency relation leads to irrelevant corrections to the temperature dependence of the correlation length. 

\begin{figure}
  \centering
  \begin{minipage}[c]{.46\linewidth}
     \includegraphics[width=0.9\columnwidth]{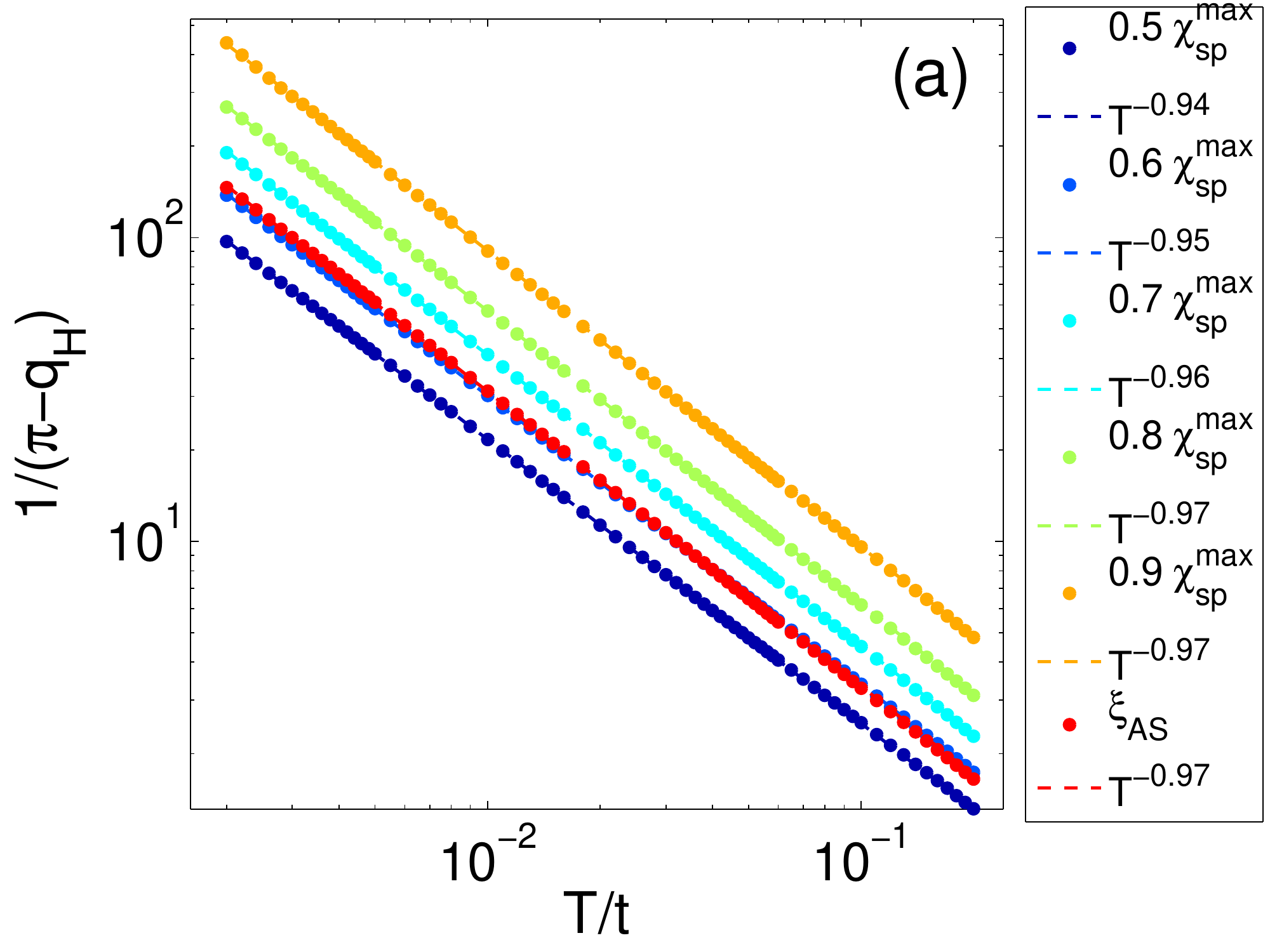}
   \end{minipage} \hfill
  \begin{minipage}[c]{.46\linewidth}
    \includegraphics[width=0.9\columnwidth]{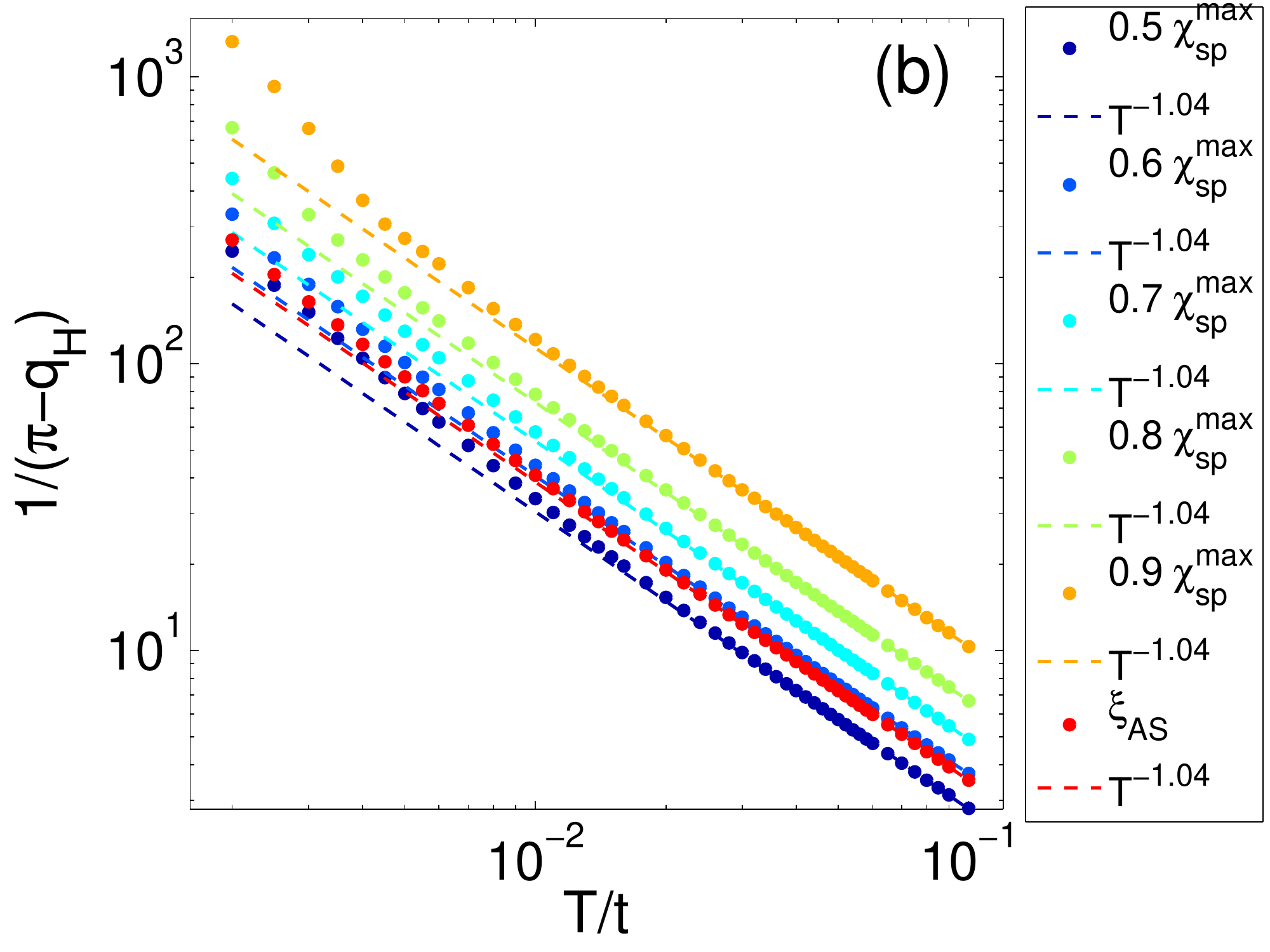}
   \end{minipage}
  \caption{Log-log plot for the temperature scaling of the correlation length. The scaling is estimated from the width measured along one of the reciprocal lattice wave vectors $(\pi-q_H)$ at various fractions of the maximum height. On (a), for $U=5.56$, the critical doping corresponds to $n_c=1.2007$, where the Fermi surface is tangent to the antiferromagnetic zone boundary. The temperature scale is too small to detect possible logarithmic corrections. The results are consistent with $z=1$. On (b), deviations from $1/T$ occur at low temperature because, for the chosen value $U=6$, the calculation is at the critical doping $1.20096$, slightly away from the point where the Fermi surface is tangent to the antiferromagnetic zone boundary. Also shown, $\xi_{AS}$ obtained from the Ornstein-Zernicke form Eqn.(\ref{xsi2}) using $U_{sp}$ from the self-consistency relation Eqn.(\ref{sum_rule})}
  \label{Fig:Chi_sp}
\end{figure}

The scaling of the correlation length can be obtained from numerical calculations by plotting,  for example, the inverse of the width of the real part of the spin susceptibility at zero frequency measured at various fractions of the maximum value. For $U=5.56$, the critical doping corresponds to $n_c=1.2007$ where the Fermi surface is tangent to the antiferromagnetic zone boundary. For that case, Fig.~\ref{Fig:Chi_sp}(a) shows that whether we measure the width at half-maximum or at some other fraction of the maximum, that width scales essentially as $1/T$, with small deviations probably coming from the fact that we have not reached the asymptotic regime. We also show on this figure the correlation length $\xi_{AS}$ obtained from the Ornstein-Zernicke form Eqn.(\ref{xsi2}) using $U_{sp}$ from the self-consistency relation Eqn.(\ref{sum_rule}). Deviations from the $1/T$ power law occur if we measure the width of the spin susceptibility too far in the tails, i.e. for a small fraction of the maximum (not shown). As demonstrated in Fig.~\ref{Fig:Chi_sp}(b), deviations from $1/T$ also occur at low temperature for a value of $U$ ($=6$ in the present example) where the critical point does not occur precisely when the Fermi surface is tangent to the antiferromagnetic zone boundary.  

To conclude this section, we show that one can easily obtain the temperature scaling for two more quantities. First, from the general form for the susceptibility Eqn.(\ref{Suscep_simple}) used above, the static susceptibility at $(\pi,\pi)$ scales as
\begin{equation}
\chi_{sp}(\mathbf{0},0)=\frac{2\xi^{2}}{U_{sp}\xi_{0}^{2}}\sim \frac{1}{\sqrt T},
\label{Chi(T)TPSC}
\end{equation}
which can be checked directly numerically, or more simply deduced from the temperature dependent results for $\xi^2$, Eqn.(\ref{Xsi_QCP}), and $\xi_0^2$, Eqn.(\ref{Xsi_0}).

\begin{figure}[ht]
  \centering
   \begin{minipage}[c]{.46\linewidth}
     \includegraphics[width=0.9\columnwidth]{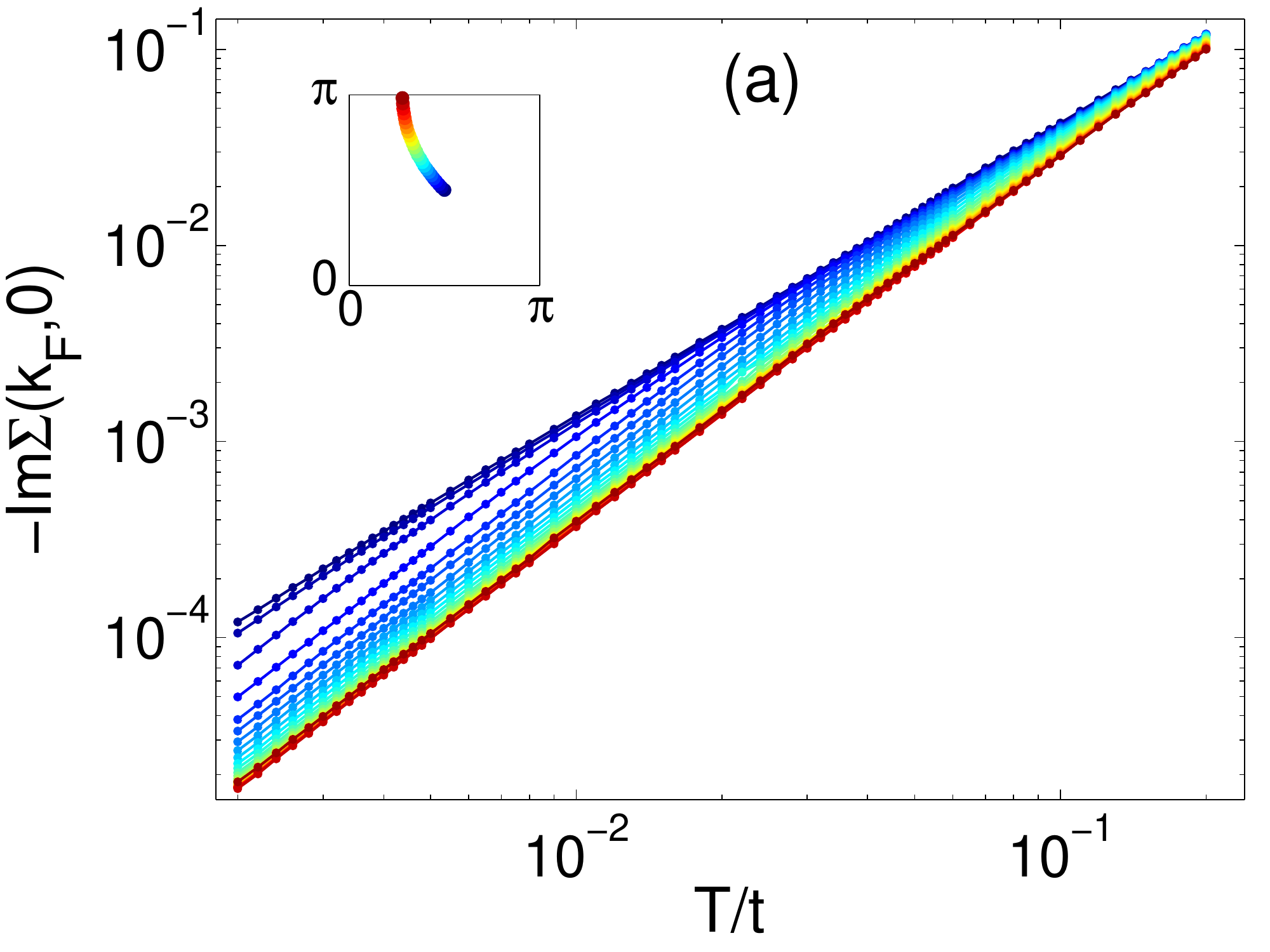}
   \end{minipage} \hfill
  \begin{minipage}[c]{.46\linewidth}
    \includegraphics[width=0.9\columnwidth]{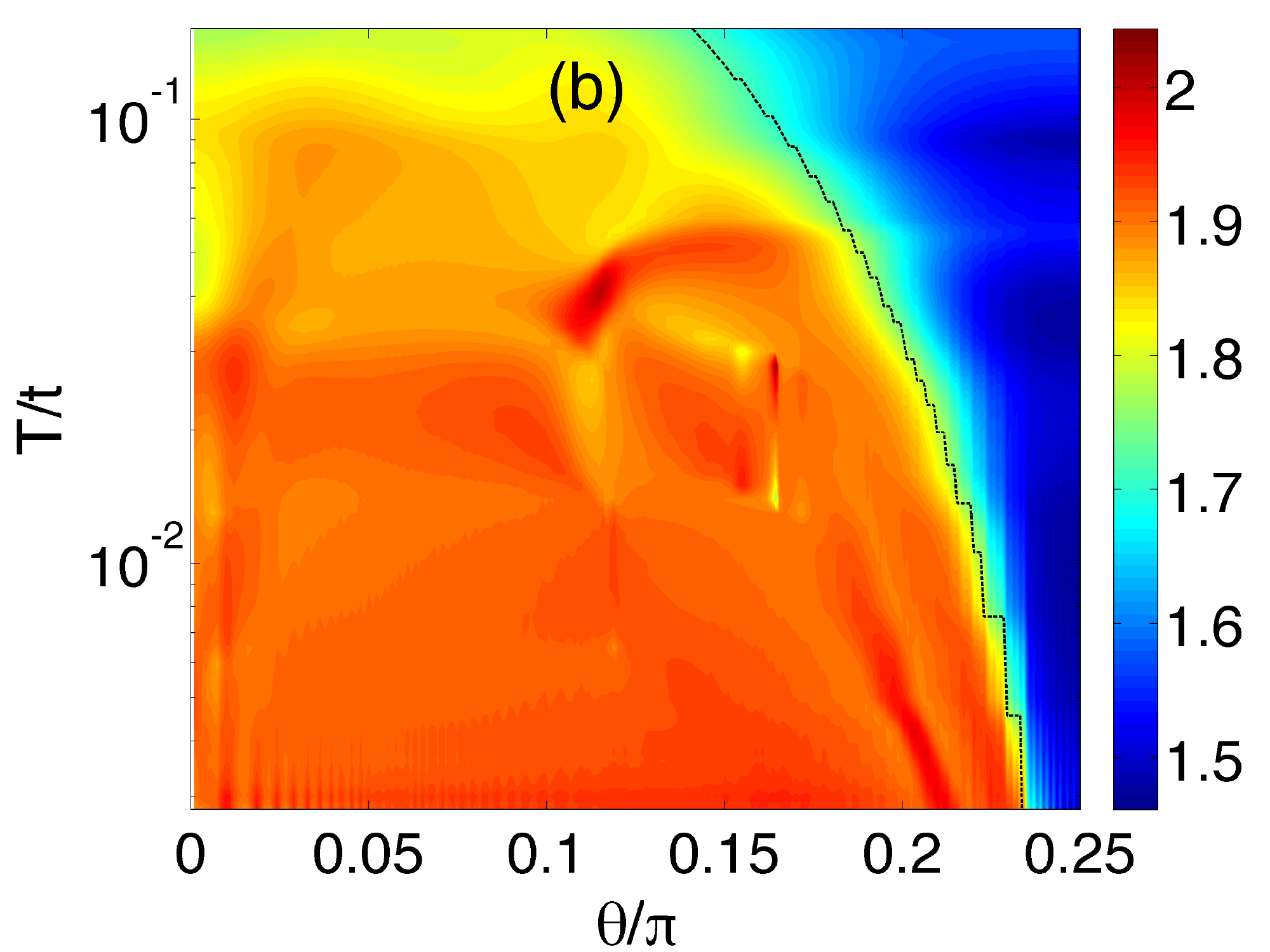}
   \end{minipage}
      \begin{minipage}[c]{.46\linewidth}
     \includegraphics[width=0.9\columnwidth]{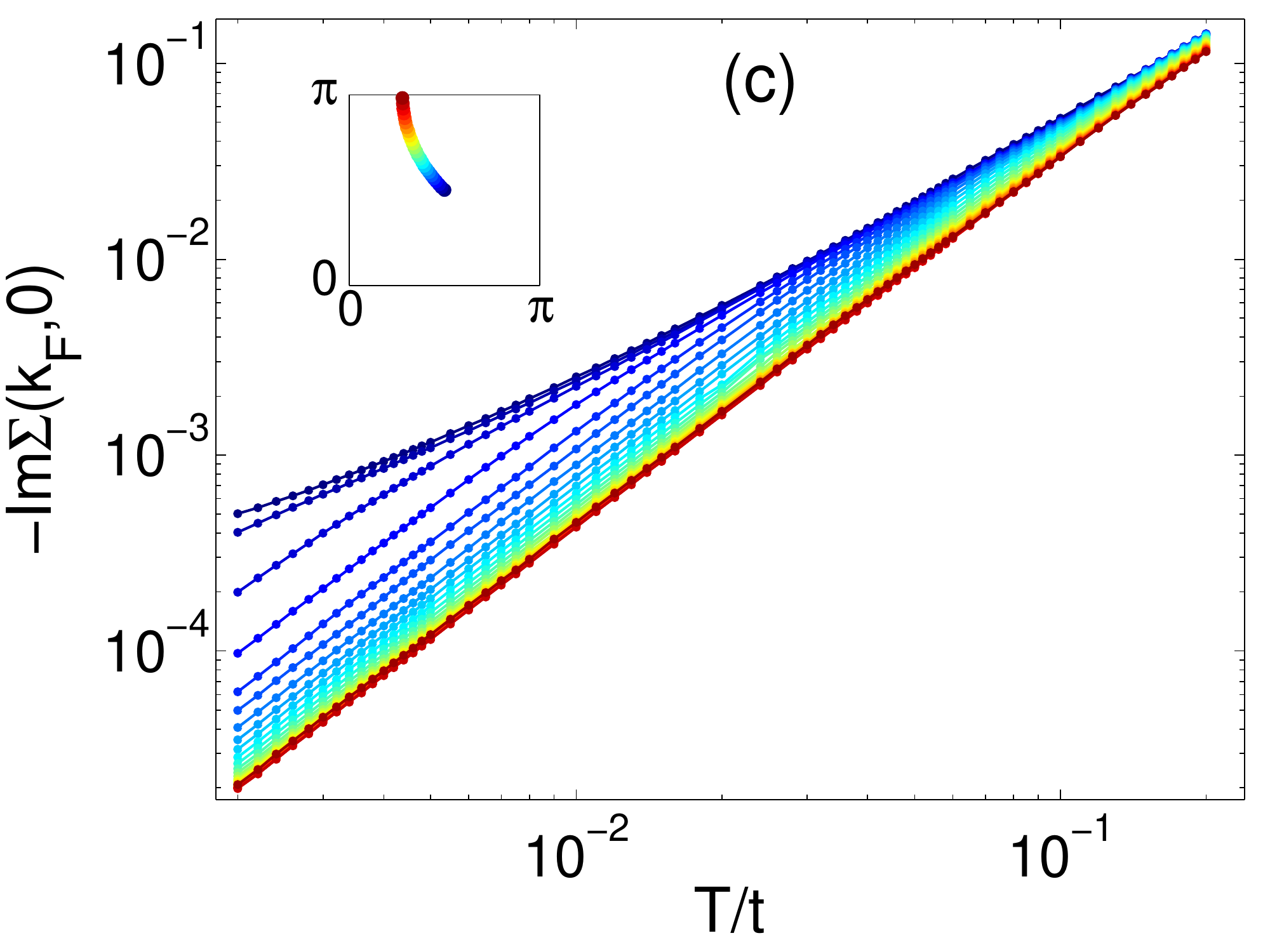}
   \end{minipage} \hfill
  \begin{minipage}[c]{.46\linewidth}
    \includegraphics[width=0.9\columnwidth]{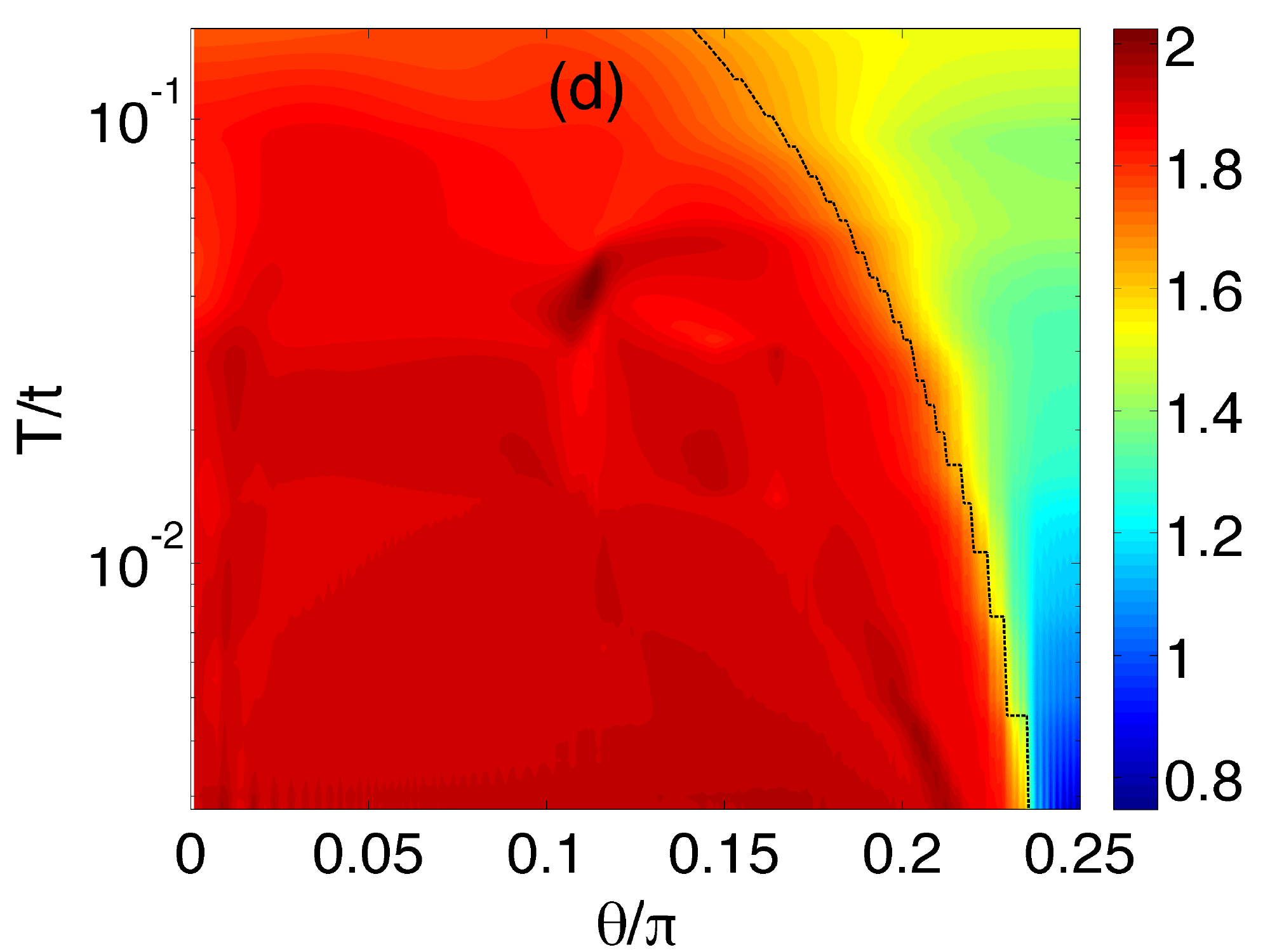}
   \end{minipage}
   
  \caption{On (a), log-log plot of the temperature dependence of the imaginary part of the self-energy at various color-coded points on the Fermi surface. The color code is in the inset. On (b), the local exponent is given as a function of angle and temperature. The points near the hot spot at $\theta=\pi/4$ behave as $T^{3/2}$ over the accessible temperature range. Calculations are done with $U=5.56$, $t^{\prime}=-0.175$ and $t^{\prime\prime}=0.05$ at the quantum critical filling $n_c=1.2007$. The lower figures are the corresponding results for $U=6$, $t^{\prime}=-0.175$ and $t^{\prime\prime}=0.05$ at the quantum critical point $n=1.20096$. Since in that case the Fermi surface is not exactly tangent to the antiferromagnetic zone boundary, the $T^{3/2}$ behavior near $\theta=\pi/4$ is recovered only if the temperature is high enough that details of the Fermi surface cannot be resolved. The black lines on (b) and (d) are the curves defined by $T_{onset}=v_F \delta k_\perp(\theta)/2$ where $\delta k_\perp(\theta)$ is the component of $\mathbf{k}_F-(\pi/2,\pi/2)$ parallel to $\mathbf{v}_F(\pi/2,\pi/2)$ at a given angle $\theta$.}
  \label{Fig:Self_T_TPSC}
\end{figure}

Finally, the nuclear magnetic resonance (NMR) relaxation rate $T_1^{-1}$ can be obtained from the two-dimensional version of the Moriya formula
\begin{equation}
T_1^{-1}=T\lim_{\omega\rightarrow 0}\int |A_\mathbf{q}|^2\frac{\chi''_{sp}(\mathbf{q},\omega)}{\omega}d^2q
\end{equation}
where $|A_{\mathbf{q}}|$ is proportional to the hyperfine matrix element. Taking this as a constant and using the general scaling form Eqn.(\ref{ChiSpScaling}), a simple change of integration variable shows that
\begin{equation}
T_1^{-1}=T\lim_{\omega\rightarrow 0}\int \frac{1}{\sqrt{T}}\frac{g(\frac{q}{T},\frac{\omega}{T})}{\omega}d^2q
\sim T^{3/2}.
\end{equation}
However, the integral over momenta $q$ also contains contributions far from the peak in the susceptibility. There the susceptibility is essentially temperature independent. There is thus a Korringa contribution $T_1^{-1}\sim T$ that is dominant at low temperature.

\subsubsection{Self-energy}%

To find the scaling of the self-energy, we use the scaling form of the susceptibility Eqn.(\ref{ChiSpScaling})
to rewrite the self-energy Eqn.(\ref{SelfWithChipp}) in
the form%
\begin{equation}
\Sigma^{\prime\prime R}\left(  \mathbf{k}_{F},\omega\right)  =-\frac{3UU_{sp}%
}{8}\frac{1}{2v_{F}}\int\frac{d^{d-1}q_{\Vert}}{\left(  2\pi\right)  ^{d-1}%
}\int\frac{d\omega^{\prime}}{\pi}\left[  n\left(  \omega^{\prime}\right)
+f\left(  \omega+\omega^{\prime}\right)  \right] \frac{1}{\sqrt{T}}g\left(\frac{q_{\Vert}}{T},\frac{(\omega+\omega')/v_F}{T};\frac{\omega}{T}\right) %
\end{equation}
Specializing to two dimensions and remembering the scaling of the Bose and Fermi functions with frequency and temperature, we change integration variables to $x=\frac{q_{\Vert}}{T}$ and $y=\frac{\omega'/v_F}{T}$ and we are left with
\begin{equation}
\Sigma^{\prime\prime R}\left(  \mathbf{k}_{F},\omega\right)=T^{3/2}S\left(\frac{\omega}{T}\right)
\label{Self-Scaling}
\end{equation}
where $S(\frac{\omega}{T})$ is a scaling function.

\begin{figure}[ht]
  \centering
     \begin{minipage}[c]{.46\linewidth}
      \includegraphics[width=0.9\columnwidth]{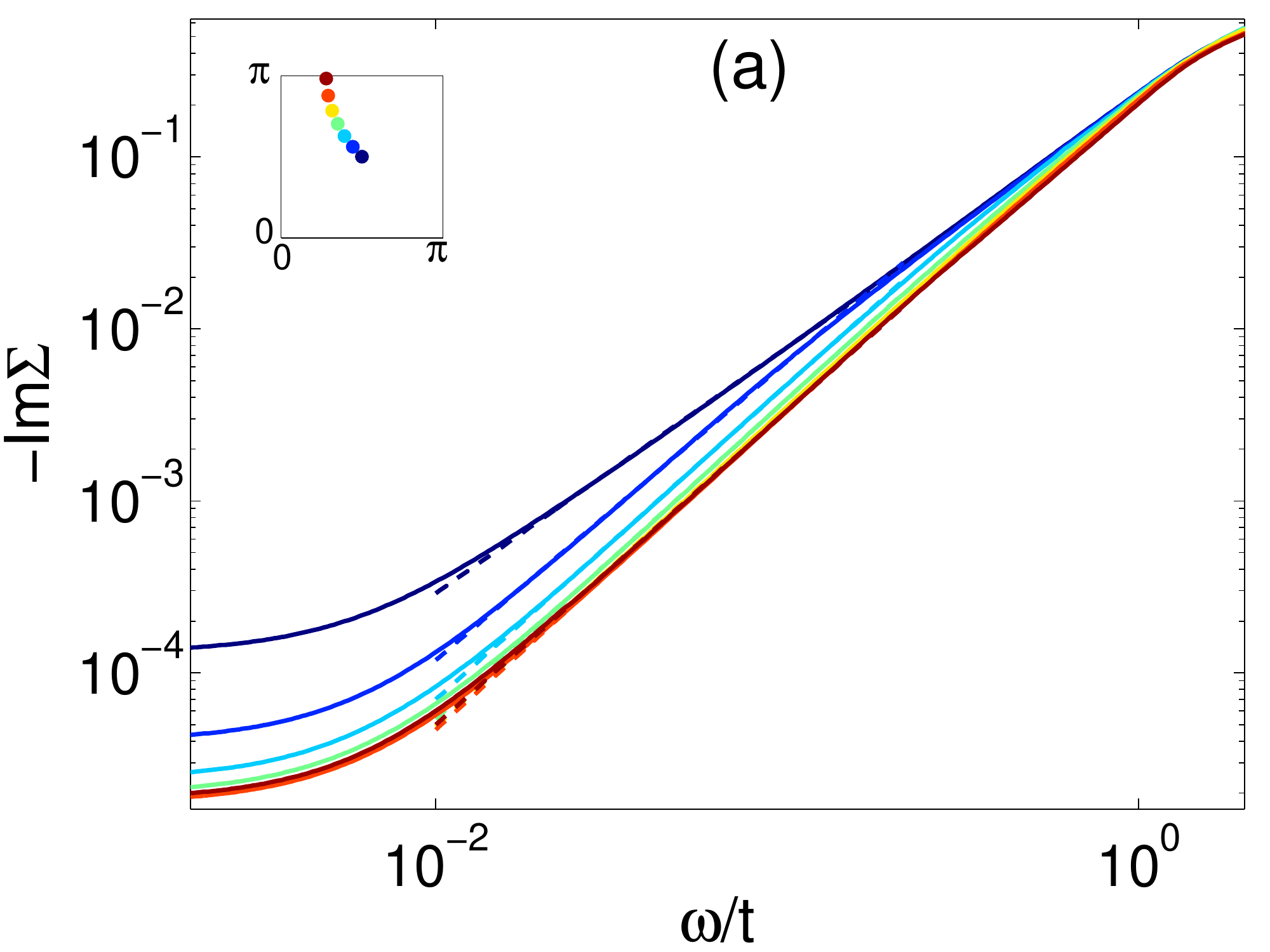}
   \end{minipage} \hfill
  \begin{minipage}[c]{.46\linewidth}
    \includegraphics[width=0.9\columnwidth]{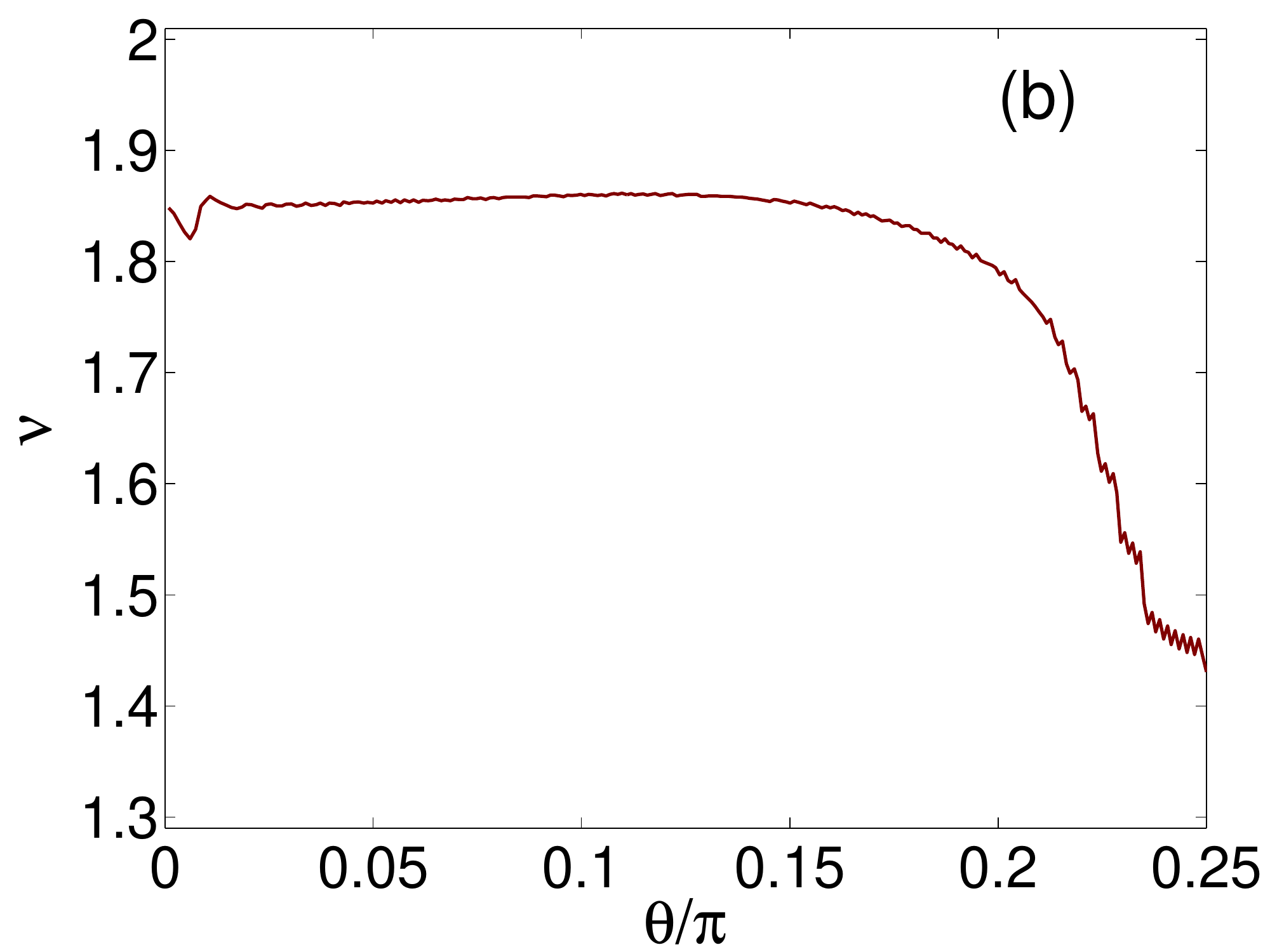}
   \end{minipage}
   \begin{minipage}[c]{.46\linewidth}
     \includegraphics[width=0.9\columnwidth]{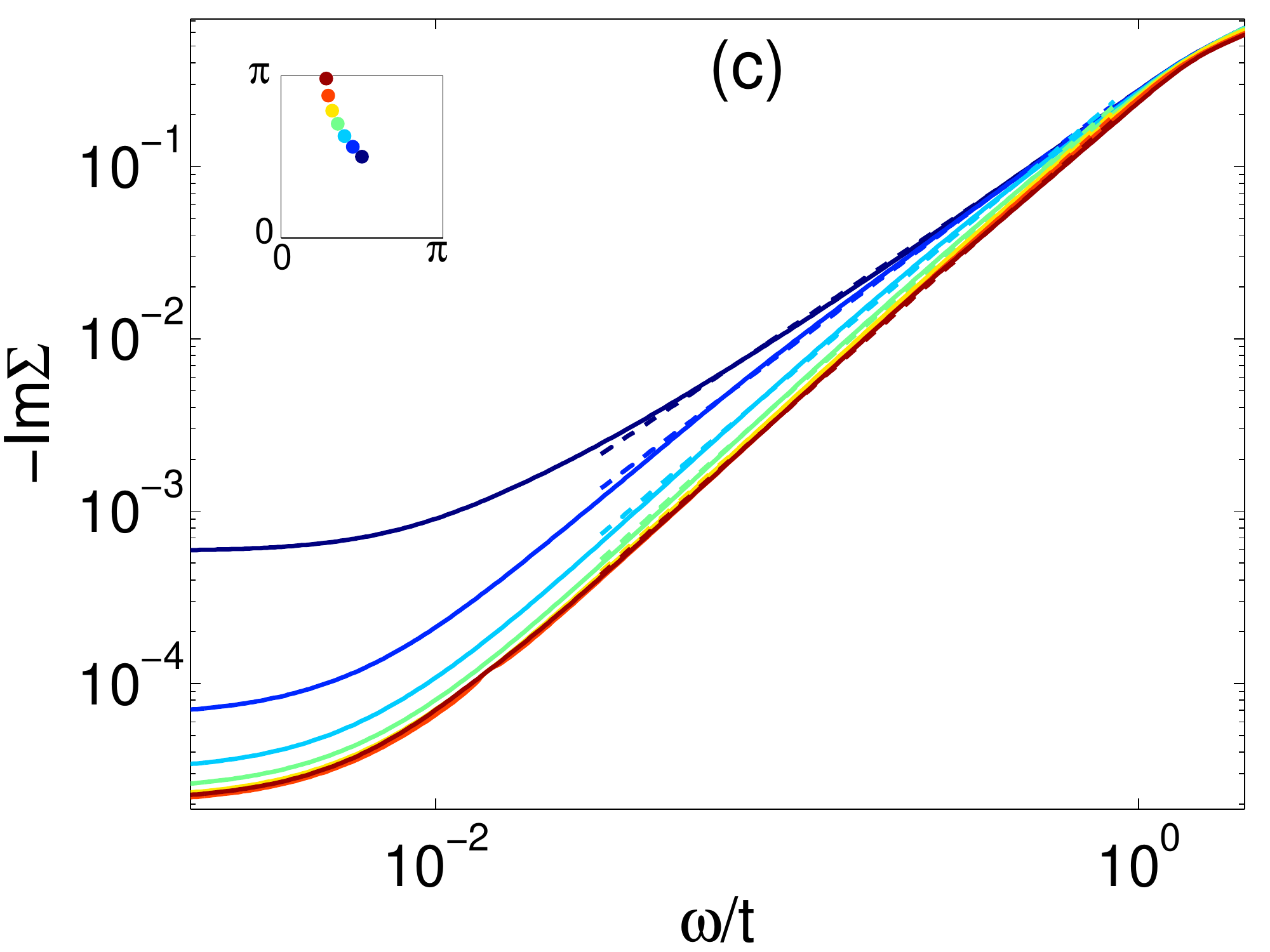}
   \end{minipage} \hfill
  \begin{minipage}[c]{.46\linewidth}
    \includegraphics[width=0.9\columnwidth]{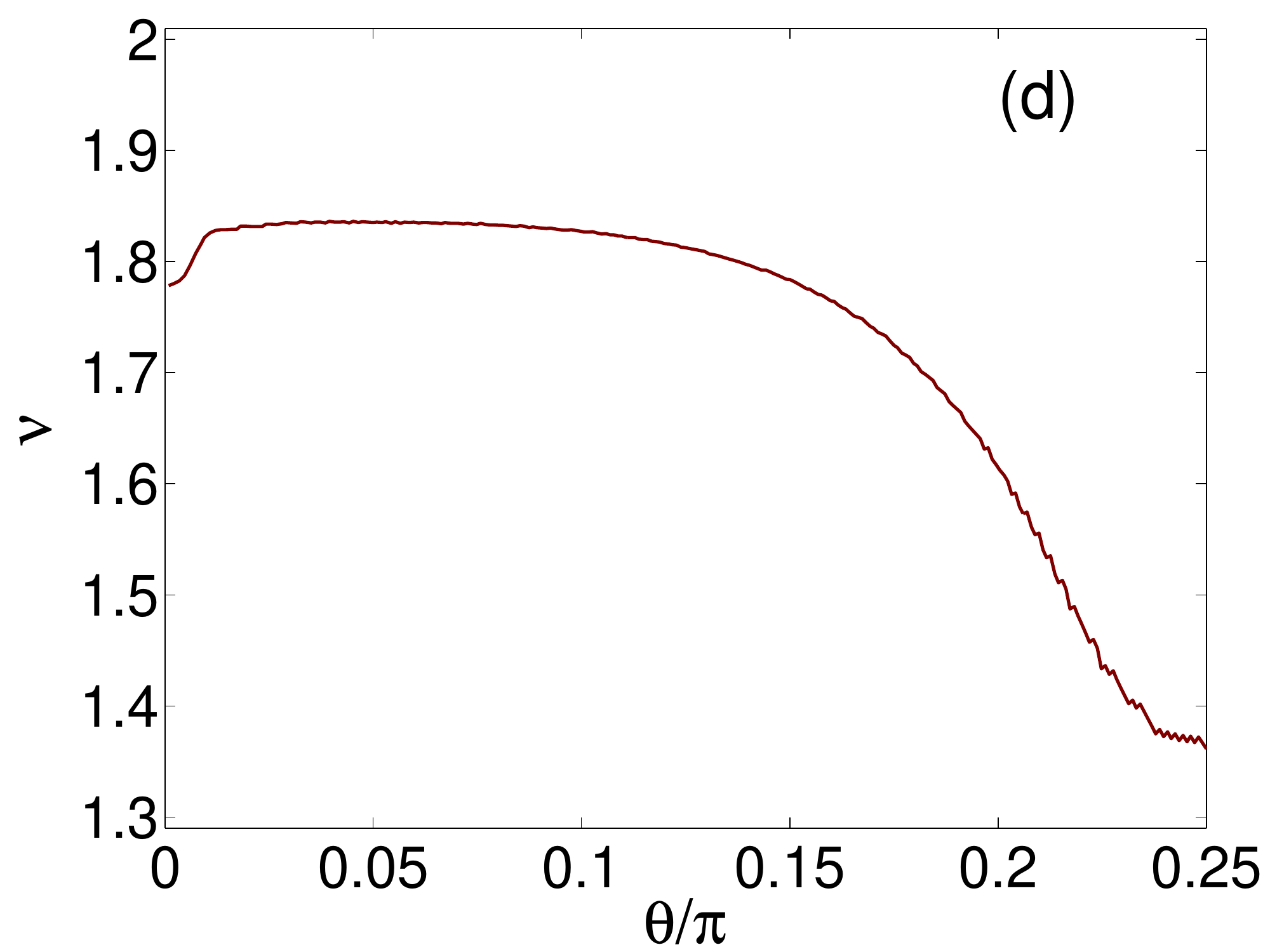}
   \end{minipage}
 
  \caption{On (a) and (c), log-log plot for the frequency dependence of $-\Sigma^{\prime\prime R}\left(  \mathbf{k}_{F},\omega\right)$ at various angles along the Fermi surface. The result of a power law fit is shown on (b) and (d). The dashed lines on (a) and (c) correspond to the fitted power laws. At the hot spot located at $\theta=\pi/4$, $-\Sigma^{\prime\prime R}$ scales as $\omega^{3/2}$. The frequency range is small because of the low temperature saturation shown on the next figure. We have verified that the crossover from $\omega^{3/2}$ to the Fermi liquid regime $\omega^{2}$ occurs on a broader angular scale when the temperature is higher, as expected from the results of Fig.~\ref{Fig:Self_T_TPSC}. Calculations are done at $T=0.002$, $t^{\prime}=-0.175$ and $t^{\prime\prime}=0.05$ at $U=5.56$ and $n=n_c=1.2007$ for (a) and (b) and $U=6$, $n=1.20096$ for (c) and (d).}
  \label{Fig:Self_w_TPSC}
\end{figure}

The latter result can be checked numerically at $\omega=0$ where we expect $\Sigma^{\prime\prime R}\left(  \mathbf{k}_{F},0\right)\sim T^{3/2}$. Figs.~\ref{Fig:Self_T_TPSC}(a) and \ref{Fig:Self_T_TPSC}(c) displays the temperature dependent scattering rate for various angles $\theta$ along the Fermi surface for $U=5.56$ and $n=n_c=1.2007$ and $U=6$ and $n=1.20096$, respectively. The line $\theta=0$ is horizontal in the Brillouin zone appearing in the inset. In Fig.~\ref{Fig:Self_T_TPSC}(a), at the hot spot located at $\theta = \pi/4$, we recover the predicted result, $T^{3/2}$. This is best illustrated in Fig.~\ref{Fig:Self_T_TPSC}(b) by a plot of the local slope of the preceding log-log plot. As we move away from the hot spot, Fermi liquid behavior appears to be recovered. There are well known logarithmic corrections in two dimensions~\cite{Hodges:1971} that may explain why we seem to deviate from exactly $T^2$. One also notices that the $T^{3/2}$ behavior occurs over a wider range of angles when the temperature is high. This is easily understood from Figs.~\ref{Fig:Self_T_TPSC}(b) and \ref{Fig:Self_T_TPSC}(d) that illustrates how temperature affects the domain where the pseudo-nesting occurs in the spin susceptibility. The solid black line in Figs.~\ref{Fig:Self_T_TPSC}(b) and \ref{Fig:Self_T_TPSC}(d) is defined by $T_{onset}=v_F \delta k_\perp(\theta)/2$ where $\delta k_\perp(\theta)$ is the component of $\mathbf{k}_F-(\pi/2,\pi/2)$ parallel to $\mathbf{v}_F(\pi/2,\pi/2)$ at a given angle $\theta$. Figs.~\ref{Fig:Self_T_TPSC}(c) and Figs.~\ref{Fig:Self_T_TPSC}(d) are for $U=6, n=1.20096$. Since at this quantum critical point the Fermi surface is not tangent to the antiferromagnetic zone boundary, the $T^{3/2}$ behavior occurs near $\theta = \pi/4$ only at high enough temperature. At low temperature, deviations become apparent.  

\begin{figure}
  \centering
  \includegraphics[width=0.4\columnwidth]{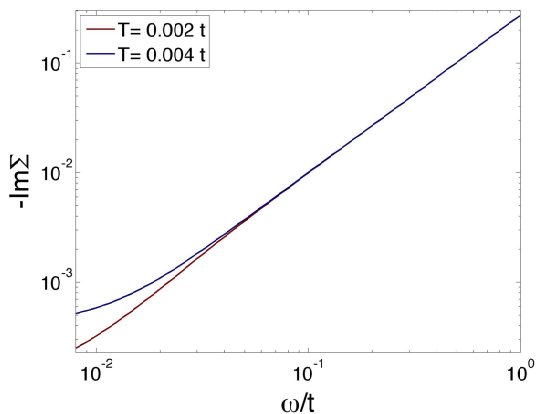}\\
  \caption{Log-log plot for the frequency dependence of $-\Sigma^{\prime\prime R}$ at the hot spot for two different temperatures. The saturation at low frequency occurs at higher frequency when the temperature is higher. Calculations are done with $U=6$, $t^{\prime}=-0.175$ and $t^{\prime\prime}=0.05$ at the quantum critical filling $n_c=1.201$ appropriate for electron-doped cuprates.}\label{Fig:Self_Saturation_TPSC}
\end{figure}

When $\omega\gg T$, the scaling form for the self-energy Eqn.(\ref{Self-Scaling}) predicts $\Sigma^{\prime\prime R}\left(  \mathbf{k}_{F},\omega\right)\sim \omega^{3/2}$. However, at zero temperature, or when $\omega\gg T$, the analytical approach taken above fails because the expansion of the spin susceptibility in $\omega/T$ and $q/T$ is no longer justified and we cannot expect the latter result to be correct. Nevertheless, TPSC can be solved numerically. To set the stage for the next section where calculations are performed analytically directly at zero temperature, we show in Fig.~\ref{Fig:Self_w_TPSC} the result of the numerical calculations for $\omega\gg T$ for two values of the interaction strength at a doping near $n_c$. On Figs.~\ref{Fig:Self_w_TPSC}(a) and \ref{Fig:Self_w_TPSC}(b), $U=5.56$ while $U=6$ on Figs.~\ref{Fig:Self_w_TPSC}(c) and \ref{Fig:Self_w_TPSC}(d). At the hot spot for $U=5.56$, the scaling of the imaginary part of the self energy is very close to the expected result $\omega^{3/2}$. For $U=6$, there is a larger discrepancy with the predicted scaling because at $n=1.20096$ the Fermi surface does not touch $(\pi/2,\pi/2)$ and thus the present theory does not apply anymore at low temperature. Away from the hot spot, Fermi liquid behavior is recovered. Again, logarithmic corrections are inaccessible from the numerical solution of the full TPSC equations because of the limited range of available frequencies: scaling is no-longer valid at frequency of the order of the Fermi energy, while at low frequency there is a saturation arising from the finite temperature. This saturation is illustrated in Fig.~\ref{Fig:Self_Saturation_TPSC}. We discuss analytically the $T=0$ regime in the following section where logarithmic corrections are found. The $1/\sqrt T$ temperature dependence of the static $(\pi,\pi)$ spin susceptibility obtained above will also be recovered.

\section{Finite frequency $T=0$ results from field theory}\label{Sec:FieldTheory}

\subsection{Lagrangian and scaling}\label{Sub:Lagrangian}

In this section we study the properties of fermionic excitations close to the hot spots within the field-theoretic framework of a spin-fermion model. This effective low-energy theory describes fermions with a parabolic dispersion (represented by fields, $\psi$) interacting with SDW fluctuations (represented by a O(3) vector field, $\vec{\phi}$). As shown in Fig.~\ref{bzf} there are four hot-spots on the Fermi surface which are connected by the SDW wave-vector ${\bf Q}=(\pi,\pi)$.
Earlier studies\cite{PKAVC06, AIM95} of the spin-fermion model in the present context did not include the umklapp processes properly and we show in the following that a correct treatment of these terms modifies the results drastically.

We start with the two patch (denoted by $s=\pm$) model \cite{MMSS10a,MMSS10b} in the rotated $(k_x,k_y)$ coordinate system. The umklapp contributions will be discussed later. In order to simplify the notation, we have rescaled our coordinates to get rid of the Fermi velocity and curvature of the Fermi-surface. The corresponding Lagrangian takes the form
\begin{eqnarray}
{\cal{L}}=\sum_{s=\pm}\psi_s^\dagger(  \partial_\tau-is\partial_x&-&\partial_y^2)\psi_s\nonumber + \\
\lambda\vec{\phi}.(\psi_+^\dagger\vec{\tau}\psi_-&+&\psi_-^\dagger\vec{\tau}\psi_+) +\nonumber \\
\frac{N}{2}(\nabla\vec{\phi})^2+\frac{Nr}{2}\vec{\phi}^2&+&\frac{Nu}{4}(\vec{\phi}^2)^2.
\label{sf}
\end{eqnarray}
Here we have promoted each fermion field to have $N$ flavors (the flavor index is suppressed). The Yukawa-coupling, $\lambda$, is chosen to be ${\cal{O}}(1)$. As a result of this, the couplings of all the bosonic terms in the last line above are scaled by a factor of $N$ as they will appear naturally upon integrating out the fermion fields. We don't include the kinetic energy of the boson, $(\partial_\tau\vec{\phi})^2$, as this is an irrelevant term \cite{MMSS10b}.

The bare fermion propagator is given by,
\begin{equation}
G_s^0(k)=\frac{1}{-i   k_\tau+sk_x+k_y^2},~~k=(k_\tau,{\bf k}).
\end{equation}
The Fermi surfaces are located at $k_x=k_y^2$ and $k_x=-k_y^2$ for patch $-$ and $+$ respectively. From Fig.~\ref{bzf} we immediately observe that ${\bf Q}-(0,2\pi)$ and also ${\bf Q}-(2\pi,0)$ connect two more points in the BZ. However, it is convenient to fold back the points within the BZ, which effectively gives rise to two more patches. These can be described by rotating the original patches by $\pi/2$. Let us denote $(k_y,k_x)$ by $\tilde {\bf k}$. Then, the equations of these two additional fermi surfaces are given by $k_y=k_x^2, k_y=-k_x^2$.

Physically, these two scattering processes are very different since in the former case, the $\vec{\phi}-$fluctuation scatters fermions that disperse strongly in the direction transverse to the Fermi surface while in the latter case, they disperse strongly in the tangential direction. This will have interesting consequences in the behavior of the electron self energy as a function of the external frequency.

The rest of this section is organized as follows: In section \ref{RPA}, we compute the RPA contributions to the SDW propagator including both direct as well as umklapp processes. We then use the dressed bosonic propagator to evaluate the fermion self-energy in section \ref{FSE} at leading order in $1/N$.

\subsection{RPA polarization}\label{Sub:PolarizationFT}
\label{RPA}
\begin{figure}
\begin{center}
\includegraphics[width=0.8\columnwidth]{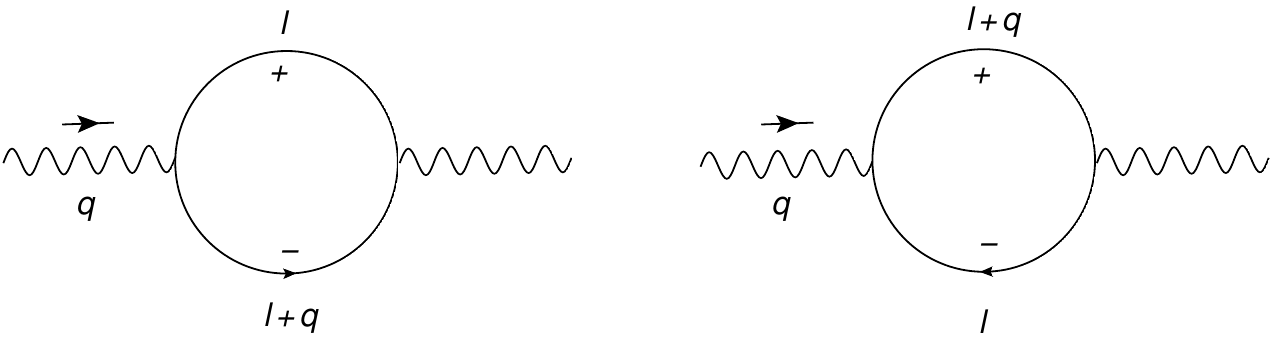}
\end{center}
\caption{The polarization bubble for the two-patch theory in Eqn.{\ref{sf}}. The internal solid lines in the loop correspond to the free fermion propagators (different patches denoted by $s=\pm$) and the external wavy lines correspond to the boson $\phi^a$.}
\label{pol}
\end{figure}

At $T=0$, the one loop polarization bubble (Fig.~\ref{pol})
for the two-patch theory
is given by,
\begin{eqnarray}
\Pi^{ab}(q)&=&2N\delta^{ab}\lambda^2\int\frac{dl_\tau d^2{\bf l}}{(2\pi)^3}[G_+^0(l)G_-^0(l+q)+G_+^0(l+q)G_-^0(l)],\\ 
\Pi(q)&=&N[\Pi_0(q_\tau,{\bf q})+\Pi_0(q_\tau,-{\bf q})],
\end{eqnarray}
where we are working with imaginary frequencies $q_\tau$ and $a,b$ denote the three SDW-polarizations. After performing the integrals, we obtain~\cite{AIM95,PKAVC06}

\begin{equation}
\Pi_0(q_\tau,{\bf q})-\Pi_0(0,0)=\frac{\lambda^2}{\sqrt{2}\pi  }\textnormal{Re}\bigg[\sqrt{E_{{\bf q}}-i  |q_\tau|}\bigg]=\frac{\lambda^2}{2\pi  }\sqrt{E_{{\bf q}}+\sqrt{E_{{\bf q}}^2+ |q_\tau|^2}}.
\label{bubble}
\end{equation}
where,
\begin{equation}
E_{{\bf q}}=\frac{q_y^2}{2}-q_x.
\label{EofQ}
\end{equation}
In the RPA propagator obtained after bubble summation, the $\Pi_0(0,0)$ contribution determines the location of the quantum critical point. It is thus convenient to add and subtract this component to make the integrals convergent. From now on, we include $\Pi_0(0,0)$ in the definition of the bubble.

To make connection with results of the previous sections, we
also quote the results for the bubble at finite temperature. In this case,
\begin{equation}
\Pi_0(q)=-\frac{\lambda^2}{4\sqrt{2}\pi}\oint_C dz\frac{1}{e^{\beta z}+1}\frac{1}{\sqrt{E_{\bf q}-i  |q_\tau|-2   z}},
\end{equation}
where the contour $C$ has to be chosen appropriately. Therefore, the above integral simplifies to,
\begin{equation}
\Pi_0(q)=-\frac{\lambda^2}{\sqrt{2}\pi}\textnormal{Re}\bigg[\int_{-\infty}^\infty dx \frac{1}{e^{\beta x}+1}\frac{1}{\sqrt{E_{\bf q}-2   x-i  |q_\tau|}} \bigg]
\end{equation}
On integrating the above equation by parts, we obtain,
\begin{eqnarray}
\Pi_0(q)&=&\frac{\lambda^2}{16\pi  ^2}\int_{-\infty}^\infty d\omega \frac{\beta}{\cosh^2[(\omega-E_{\bf q})/4   T]}\sqrt{\omega+\sqrt{\omega^2+ |q_\tau|^2}}, \\
\Pi_0(q)&=&\sqrt{T} f\bigg(\frac{|q_\tau|}{T},\frac{E_{\bf q}}{T}\bigg),
\label{ScalingPi0T}
\end{eqnarray}
where,
\begin{equation}
f\bigg(\frac{|q_\tau|}{T},\frac{E_{\bf q}}{T}\bigg)=\frac{\lambda^2}{16\pi  ^2}\int_{-\infty}^\infty dy \frac{1}{\cosh^2[(y-E_{\bf q}/T)/4  ]}\sqrt{y+\sqrt{y^2+\frac{|q_\tau|^2}{T^2}}}
\end{equation}
as found in Ref.~\onlinecite{PKAVC06}.

The RPA propagator for the SDW fluctuations is then given by,
\begin{equation}
D({\bf q},q_\tau)=\frac{1}{N}\frac{1}{{\bf q}^2+r+[\Pi_0(q_\tau,{\bf q})+\Pi_0(q_\tau,-{\bf q})+\Pi_0(q_\tau,{\bf \tilde{q}})+\Pi_0(q_\tau,-{\bf \tilde{q}})]},
\label{renprop}
\end{equation}
where we have included the RPA contribution arising from
all four hot spots on the Fermi surface and ${\bf \tilde{q}}=(q_y,q_x)$. The terms ($\Pi_0(q_\tau,{\bf q})+\Pi_0(q_\tau,-{\bf q})$) are not equal to ($\Pi_0(q_\tau,{\bf \tilde{q}})+\Pi_0(q_\tau,-{\bf \tilde{q}})$) as was incorrectly assumed by the authors of Ref.~\onlinecite{PKAVC06}. At the quantum critical filling $r=0$, zero Matsubara frequency $q_\tau=0$  but finite temperature, the ${\bf q^2}$ term is negligible compared to the contribution of the bubbles $\Pi_0$. Using the scaling form Eqn.(\ref{ScalingPi0T}) and keeping only terms linear in $q$ in $E_{\bf q}$ (Eqn.~\ref{EofQ}), one recovers the zero frequency limit of the previous result Eqn.(\ref{ChiSpScaling}) for the spin susceptibility. Naively doing the analytical continuation in frequency, the full scaling form would also follow. For the rest of the computations, we consider $T=0$ and carefully take into account logarithmic corrections that were beyond the reach of the previous calculation.

Before we proceed to evaluate the fermionic self-energy, let us compute the forms of the real and imaginary parts of the retarded polarization bubble at $T=0$, $\Pi_R({\bf q},\Omega)$, where $i|q_\tau|\rightarrow\Omega+i0^+$. For the imaginary part, we obtain from Eqn.~(\ref{bubble}),
\begin{eqnarray}
\textnormal{Im}\Pi_R({\bf q},\Omega)&=&\frac{\lambda^2}{2\sqrt{2}\pi  }\textnormal{Im}\bigg[\sqrt{E_{{\bf q}}-  \Omega-i0^+}+\sqrt{E_{{\bf q}}+  \Omega+i0^+}\bigg], \\
\textnormal{Im}\Pi_R({\bf q},\Omega)&=&\frac{-\lambda^2}{2\sqrt{2}\pi  }\bigg[\theta(  \Omega-E_{{\bf q}})\sqrt{  \Omega-E_{{\bf q}}}-\theta(-  \Omega-E_{{\bf q}})\sqrt{-  \Omega-E_{{\bf q}}}\bigg],
\end{eqnarray}
which is chosen in a way such that $\Omega\textnormal{Im}\Pi_R({\bf q},\Omega)<0$.
The real part can also be obtained from Eqn.~(\ref{bubble}) or from the Kramers-Kronig relation
\begin{eqnarray}
\textnormal{Re}\Pi_R({\bf q},\Omega)=\frac{\lambda^2}{2\sqrt{2}\pi  }\bigg[\theta(E_{{\bf q}}-  \Omega)\sqrt{E_{{\bf q}}-  \Omega}+\theta(  \Omega+E_{{\bf q}})\sqrt{  \Omega+E_{{\bf q}}}\bigg].
\end{eqnarray}
These results agree with those of Refs.~\onlinecite{AIM95,PKAVC06}. The $\Omega\ll E_{\bf q}$ limit calculated in Ref.~\onlinecite{Benard:1993} also agrees with the above.
\subsection{Electron self-energy}\label{Sub:SelfFT}
\label{FSE}
\begin{figure}
\begin{center}
\includegraphics[width=0.3\columnwidth]{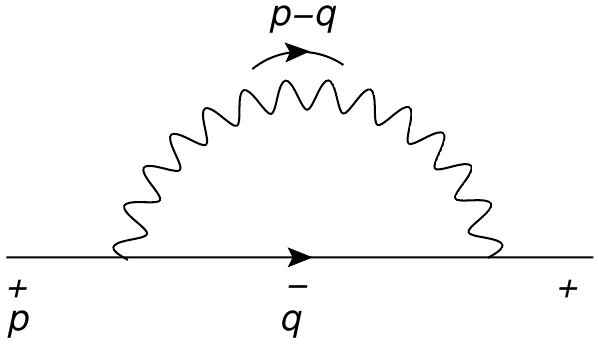}
\end{center}
\caption{The electron self energy $\Sigma_+(p)$ at leading order in $1/N$. The $\vec{\phi}-$ propagator includes the one loop bubbles evaluated earlier.}
\label{se}
\end{figure}

We are interested in evaluating the electron self energy (Fig.~\ref{se}) at $T=0$, $\Sigma(p)$, which at leading order in $1/N$ is given by,
\begin{equation}
\Sigma_\pm(\k, i\omega_n) =\frac{3\lambda^2}{\beta} \sum_{\Omega_n} \int \frac{d^2 q}{(2 \pi)^2} \, G_\mp(\k-\q,i \omega_n-i \Omega_n) D(\q, i\Omega_n)
\end{equation}

After analytic continuation $i\omega_n \to \omega + i 0^+$ we obtain the following expression for the imaginary part of the retarded self-energy
\begin{equation}
\text{Im} \Sigma^R_\pm(\k,\omega) =3\lambda^2\int  \frac{d^2 q}{(2 \pi)^2} \, \big[ n_F(-\xi^\mp_{\k-\q})+n_B(\omega-\xi^\mp_{\k-\q})\big] \text{Im} D_R(\q,\omega-\xi^\mp_{\k-\q}),
\end{equation}
where $\xi^+_\k = k_x+k_y^2$ and $\xi^-_\k = -k_x+k_y^2$.

At $T=0$ and $\k=0$ we are left with
\begin{eqnarray}
\text{Im} \Sigma^R_+(0,\omega) &=& 3\lambda^2\int  \frac{d^2 q}{(2 \pi)^2} \, \big[ \Theta(\xi^-_{-\q})-\Theta(\xi^-_{-\q}-  \omega)\big] \text{Im} D_R(\q,\omega-\xi^-_{-\q}) \notag \\
&=&  \frac{3\lambda^2}{4\pi^2  } \int dq_y \int_{-q_y^2}^{-q_y^2+  \omega} dq_x \, \text{Im} D_R(\q,\omega-\xi^-_{-\q}) \notag \\
&\overset{\omega\rightarrow0}{\approx}& \frac{3\lambda^2 \omega}{4\pi^2} \int dq_y  \, \text{Im} D_R(\q,\omega-\xi^-_{-\q}) \Big|_{q_x \to -q_y^2}
\label{eq1}
\end{eqnarray}
Note that $\xi_{-{\bf q}}^-=q_x+q_y^2\equiv0$ for $q_x \to -q_y^2$. In terms of $\Pi_R$, this can be rewritten as,
\begin{equation}
\text{Im} \Sigma^R_+(0,\omega) = \frac{3\lambda^2\omega}{4\pi^2N} \int dq_y \frac{-\text{Im} \Pi_R^{tot}(\q,\omega)}{(q_y^2+\text{Re} \Pi_R^{tot}(\q,\omega))^2+(\text{Im} \Pi_R^{tot}(\q,\omega))^2},
\end{equation}
where $\Pi_R^{tot}(\q,\omega)$ is the total retarded RPA bubble including direct and umklapp terms and we have ignored the $q_x^2\sim q_y^4$ term in the denominator.

For $q_x=-q_y^2$, we get $E_{\bf q}=3q_y^2/2$, $E_{-\bf q}=-q_y^2/2$, $E_{\bf \tilde{q}}=q_y^4/2-q_y$ and $E_{-\bf \tilde{q}}=q_y^4/2+q_y$. Therefore, Re(Im)$\Pi_R^{tot}$ are given by,
\begin{eqnarray}
\text{Re} \Pi_R^{tot}(\q,\omega)=\frac{\lambda^2}{4\pi  }\bigg[\sqrt{3q_y^2-2    \omega} \, \theta(3q_y^2-2    \omega)&+&\sqrt{3q_y^2+2    \omega} \, \theta(3q_y^2+2    \omega) \nonumber\\
+\sqrt{-2    \omega-q_y^2} \, \theta(-2    \omega-q_y^2)&+&\sqrt{2    \omega-q_y^2} \, \theta(2    \omega-q_y^2) \nonumber\\
+\sqrt{q_y^4-2q_y-2    \omega}\,\theta(q_y^4-2q_y-2   \omega)&+&\sqrt{q_y^4-2q_y+2   \omega} \, \theta(q_y^4-2q_y+2   \omega) \nonumber\\
+\sqrt{q_y^4+2q_y-2    \omega} \, \theta(q_y^4+2q_y-2    \omega)&+&\sqrt{q_y^4+2q_y+2    \omega}\theta(q_y^4+2q_y+2    \omega) \bigg]\nonumber\\
\text{Im} \Pi_R^{tot}(\q,\omega)=-\frac{\lambda^2}{4\pi   }\bigg[\sqrt{2    \omega-3q_y^2} \, \theta(2   \omega-3q_y^2)&+&\sqrt{2   \omega+q_y^2} \, \theta(2   \omega+q_y^2)\nonumber \\
-\sqrt{-3q_y^2-2   \omega} \, \theta(-3q_y^2-2    \omega)&-&\sqrt{q_y^2-2   \omega} \, \theta(q_y^2-2   \omega) \nonumber\\
+\sqrt{2    \omega+2q_y-q_y^4} \, \theta(2    \omega+2q_y-q_y^4)&+&\sqrt{2    \omega-2q_y-q_y^4} \, \theta(2    \omega-2q_y-q_y^4)\nonumber\\
-\sqrt{2q_y-q_y^4-2    \omega} \, \theta(2q_y-q_y^4-2    \omega)&-&\sqrt{-2q_y-q_y^4-2    \omega} \, \theta(-2q_y-q_y^4-2    \omega) \bigg]\nonumber\\
\end{eqnarray}
As a starting point, we can drop the $q_y^4$ terms in the limit of small $\omega$ and retain only the $q_y$ terms in the
$\tilde{{\bf q}}$ contributions that we take into account.
This is a consistent way of handling these terms, since if typical $q_y\sim\omega$, then $q_y^4$ is smaller than $q_y$, so that it is justified to drop these terms. Since the integrand is an even function of $q_y$, we integrate only over $q_y>0$. Then, the expression for self energy reduces to,

\begin{eqnarray}
&=&\frac{3\lambda^2\omega}{4\pi^2N}\int_{-\infty}^\infty dq_y\frac{-\text{Im} \Pi_R^{tot}(\q,\omega)}{(q_y^2+\text{Re} \Pi_R^{tot}(\q,\omega))^2+(\text{Im} \Pi_R^{tot}(\q,\omega))^2}\nonumber\\
&\approx&\frac{3    \omega}{8\pi^3N}\bigg[\int_{0}^{   \omega} dq_y A(q_y,  \omega)+\int_{  \omega}^{\sqrt{2  \omega/3}}dq_y B(q_y,  \omega)+\int_{\sqrt{2  \omega/3}}^{\sqrt{2  \omega}}dq_y C(q_y,  \omega)+\int_{\sqrt{2  \omega}}^{\infty}dq_y D(q_y,  \omega)\bigg]\nonumber\\
A&=&\frac{\sqrt{2    \omega+2q_y}+\sqrt{2    \omega-2q_y}+\sqrt{2    \omega-3q_y^2}+\sqrt{2    \omega+q_y^2}}{\bigg[\frac{1}{\lambda^2} q_y^2+\frac{\sqrt{2q_y+2    \omega}+\sqrt{2    \omega-2q_y}+\sqrt{3q_y^2+2    \omega}+\sqrt{2    \omega-q_y^2}}{4\pi}\bigg]^2+\frac{[\sqrt{2    \omega+2q_y}+\sqrt{2    \omega-2q_y}+\sqrt{2    \omega-3q_y^2}+\sqrt{2    \omega+q_y^2}]^2}{16\pi^2}} \nonumber\\
B&=&\frac{\sqrt{2    \omega+2q_y}-\sqrt{2q_y-2    \omega}+\sqrt{2    \omega-3q_y^2}+\sqrt{2    \omega+q_y^2}}{\bigg[\frac{1}{\lambda^2} q_y^2+\frac{\sqrt{2q_y+2    \omega}+\sqrt{2q_y-2    \omega}+\sqrt{3q_y^2+2    \omega}+\sqrt{2  \omega-q_y^2}}{4\pi}\bigg]^2+\frac{[\sqrt{2    \omega+2q_y}-\sqrt{2q_y-2    \omega}+\sqrt{2    \omega-3q_y^2}+\sqrt{2    \omega+q_y^2}]^2}{16\pi^2}}  \nonumber\\
C&=&\frac{\sqrt{2    \omega+2q_y}-\sqrt{2q_y-2    \omega}+\sqrt{2    \omega+q_y^2}}{\bigg[\frac{1}{\lambda^2} q_y^2+\frac{\sqrt{2q_y+2    \omega}+\sqrt{2q_y-2    \omega}+\sqrt{3q_y^2+2    \omega}+\sqrt{3q_y^2-2    \omega}+\sqrt{2  \omega-q_y^2}}{4\pi}\bigg]^2+\frac{[\sqrt{2    \omega+2q_y}-\sqrt{2q_y-2    \omega}+\sqrt{2    \omega+q_y^2}]^2}{16\pi^2}}  \nonumber\\
D&=&\frac{\sqrt{2q_y+2    \omega}-\sqrt{2q_y-2    \omega}-\sqrt{q_y^2-2    \omega}+\sqrt{ q_y^2+2    \omega}}{\bigg[\frac{1}{\lambda^2} q_y^2+\frac{\sqrt{2q_y+2    \omega}+\sqrt{2q_y-2    \omega}+\sqrt{3q_y^2+2    \omega}+\sqrt{3q_y^2-2    \omega}}{4\pi}\bigg]^2+\frac{[\sqrt{2q_y+2    \omega}-\sqrt{2q_y-2    \omega}-\sqrt{q_y^2-2    \omega}+\sqrt{q_y^2+2    \omega}]^2}{16\pi^2}}\nonumber\\
\label{reg}
\end{eqnarray}
At small frequencies the dominant contribution to the imaginary part of the self-energy comes from the second integral between $  \omega$ and $\sqrt{2  \omega/3}$, which scales as $\sim - \omega^{3/2} \log \omega$. The contributions from the other regions scale as $\sim \omega^{3/2}$ and thus are negligible at low frequencies. The correct prefactor can be obtained by expanding the numerator and the denominator of integrand $B$ for small frequencies $\omega$ and retaining only the largest terms, which gives
\begin{eqnarray}
B(q_y,  \omega)=\frac{4\pi^2\sqrt{2  \omega}}{q_y}.
\end{eqnarray}
One then integrates over $q_y$ to obtain,
\begin{equation}
\text{Im} \Sigma^R_+(0,\omega) \approx-\frac{3}{2\sqrt{2}\pi N}\omega^{3/2}\log(  \omega).
\label{finalresult}
\end{equation}
We note here that the self-energy is less singular compared to earlier works where umklapp scattering was not taken into account, in which case the self-energy scales as $\sim - \omega \log \omega$.\cite{AIM95, PKAVC06}.
\begin{figure}
\begin{center}
\includegraphics[width=.95\columnwidth]{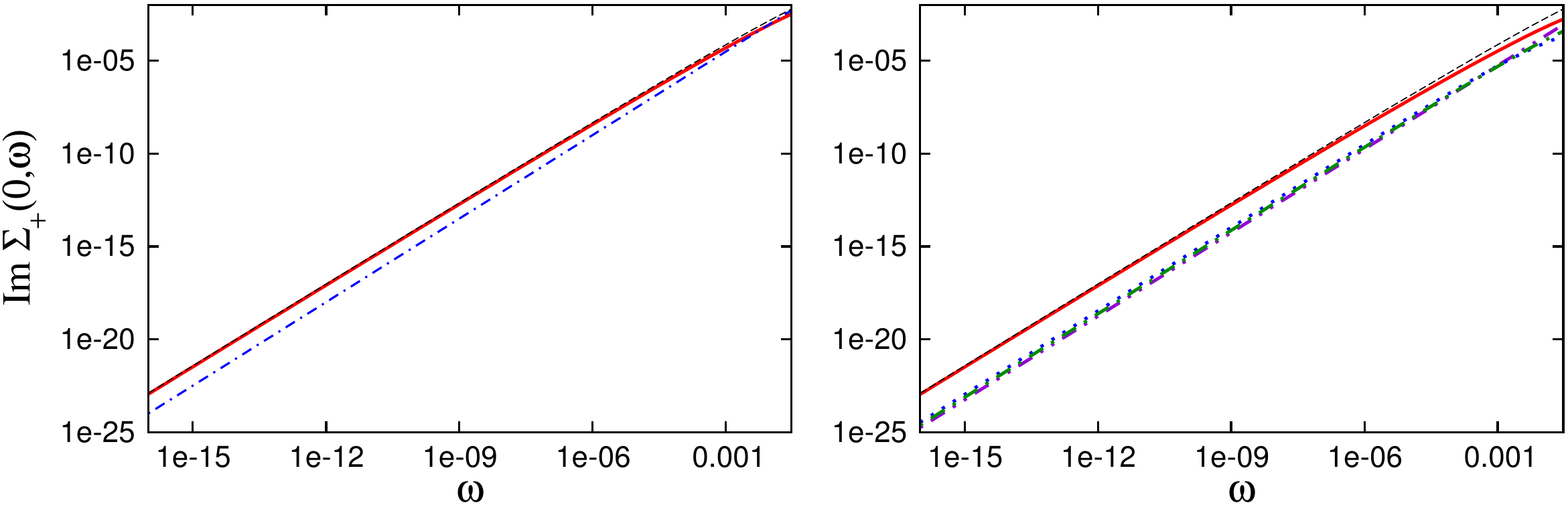}
\end{center}
\caption{Left: $\text{Im} \Sigma_R(0,\omega)$, calculated numerically from the second line of Eqn.~\eqref{eq1}, shown as a red solid line. The black dashed line is the asymptotic result Eqn.~\eqref{finalresult}. For comparison, the blue dash-dotted line indicates $\omega^{3/2}$. Right: different contributions to $\text{Im} \Sigma_R(0,\omega)$. The red line shows the dominant contribution, denoted by $B$ in the main text. Black dashed line: asymptotic result  Eqn.~\eqref{finalresult}. The other three curves show the sub-leading contributions $A, C$ and $D$, which scale as $\sim \omega^{3/2}$.}
\label{fig2}
\end{figure}

So far we haven't addressed an important issue: what happens if we include the renormalization of the boson-fermion vertex? In the two-patch theory originally considered by Altschuler et al. \cite{AIM95}, the one-loop correction to this vertex was found to be logarithmically singular. However in Appendix \ref{appver}, we show that the full four-patch theory does not have this singularity.

Based on our discussion so far, we see that the additional umklapp terms considered in our work play a very crucial role at the critical point. In the absence of these contributions, the self-energy was more singular than what we have found here. Moreover, the vertex correction was also found to be singular. However, here we have shown that the singular behavior is washed out when we include the additional scattering contributions.

\subsection{Effect of the self-consistency}\label{Sub:Scaling}

In the TPSC approach we imposed two-particle self-consistency in the form of a sum-rule that is similar to the spherical model. In the present field-theory approach, this amounts to imposing $<\vec{\phi}^2>=1$ where the expectation value is taken with respect to the fermions and bosons. We have argued that the $z=1$ scaling does not come from the self-consistency condition. To confirm this result, in this subsection we obtain the scaling of the quartic term in the boson Lagrangian. 

Integrating out the fermions, the polarization operator $\Pi_0(q_{\tau},{\bf q})$ appears in the quadratic term of the boson Lagrangian. Emphasizing the scaling only, this term is symbolically written as
\begin{equation}
\int^\Lambda (d^2q d\omega) \vec{\phi}^2 \sqrt(\omega,q).
\end{equation}
This is the most relevant quadratic term. Integrating out the large wave number modes for $q>\Lambda/s$ and rescaling $q$ and $\omega$ such that $q'=qs$, $\omega'=\omega s$ with $s>1$, returns the new cutoff $\Lambda/s$ to its original value $\Lambda$. Invariance of the quadratic term written in terms of the prime variables then imposes that $\phi=\phi' s^{\varphi}=\phi' s^{7/4}$. The effect of this tree level scaling on the quartic term is that
\begin{equation}
u\int^\Lambda (d^2q d\omega)^3 \vec{\phi}^4 \rightarrow us^{-9}s^{4\varphi}\int^\Lambda (d^2q' d\omega')^3 \vec{\phi'}^4.
\end{equation}
This in turn means that $u'=us^{-9}s^{4\varphi}=us^{-2}$ scales to zero and is thus irrelevant.

\section{Summary}\label{Sec:Summary}

We have argued that for bare interaction strengths $U$ in the intermediate coupling range, commensurate SDW fluctuations at $(\pi,\pi)$ and band parameters similar to those of electron-doped cuprates, the antiferromagnetic quantum critical point naturally occurs close to the filling where the Fermi surface points joined by $(\pi,\pi)$ are nearly tangent to each other. As long as the temperature or frequency are not too low, the limiting case of tangent Fermi surfaces  describes the physics. In this pseudo-nesting situation, the Fermi liquid behavior breaks down. Quasiparticles still exist but the self-energy and spin susceptibility, for example, are different from those predicted by Fermi liquid theory. 

We considered this problem at zero temperature, or for frequencies larger than temperature, using a field-theoretical model of gapless collective bosonic modes (SDW fluctuations) interacting with fermions. The imaginary part of the retarded fermionic self-energy close to the hot spots scales as $-\omega^{3/2}\log\omega$. This is less singular than earlier predictions of the form $-\omega\log\omega$. The difference arises from the effects of umklapp terms that were not included in previous studies.

At finite temperature, we have used TPSC to study this problem and have obtained numerical results for the one-band Hubbard model with band parameters and interaction strength appropriate for electron-doped cuprates. Neglecting logarithmic corrections, we found analytically and numerically that the correlation length $\xi$ scales like $1/T$, namely $z=1$ instead of the naive $z=2$. The static spin susceptibility $\chi$ scales like $1/\sqrt T$, and the correction $T_1^{-1}\sim T^{3/2}$ to the Korringa NMR relaxation rate is subdominant. NMR experiments are difficult in electron-doped cuprates. We also found that the imaginary part of the self-energy at the hot spot scales like $T^{3/2}$. The latter result and the $-\omega^{3/2}\log\omega$ frequency dependence of the self-energy should be experimentally verifiable with angle-resolved photoemission spectroscopy (ARPES) in electron-doped cuprates. Recent transport measurements in these compounds~\cite{Jin:2011} have found a $T^{3/2}$ behavior of the resistivity above the quantum critical point at the end of the overdoped side of the superconducting dome. While there may be a relation with the above result if antiferromagnetic fluctuations disappear at the same time, one must also be careful not to equate scattering rate with resistivity because in a simple picture it is the inverse of the scattering rates that are averaged over the Fermi surface. This suggests that in the resistivity, Fermi liquid behavior of the cold spots should short-circuit the non-Fermi liquid behavior of the hot spots.~\cite{Hlubina:1995}

\begin{acknowledgments}
We thank Erez Berg, Andrey Chubukov, Lev Ioffe and Max Metlitski for useful discussions. This research was supported by the National Science Foundation
under grant DMR-1103860 (D.C., S.S.), by the Austrian Science Fund (FWF)-Erwin Schr\"{o}dinger Fellowship J 3077-N16 (M.P.), by a MURI grant from AFOSR (S.S.) and by NSERC, the Tier I Canada Research Chair Program (A.-M. S. T.). A.-M.S.T is grateful to the Harvard Physics Department, CIFAR, and the center for materials theory at Rutgers University for support during the writing of this work. Partial support was also provided
by the MIT-Harvard Center for Ultracold Atoms. Computer intensive calculations were performed on
computers provided by CFI, MELS, Calcul Qu\'{e}bec and Compute Canada.

\end{acknowledgments}

\appendix

\section{Vertex correction}\label{Sec:Vertex}
\label{appver}
In this appendix, we compute the 1-loop correction to the Boson-Fermion vertex, which is defined as,
\begin{equation}
-\langle\psi_{-\sigma}(q)\psi_{+\sigma'}^\dagger(p)\phi^a(q)\rangle=\tau^a_{\sigma\sigma'}\Gamma_{\phi\psi_-\psi_+^\dagger}(p,q)(2\pi)^3\delta^3(q-p-q),
\end{equation}
where we are working again with imaginary frequencies.
\begin{figure}
\begin{center}
\includegraphics[width=0.3\columnwidth]{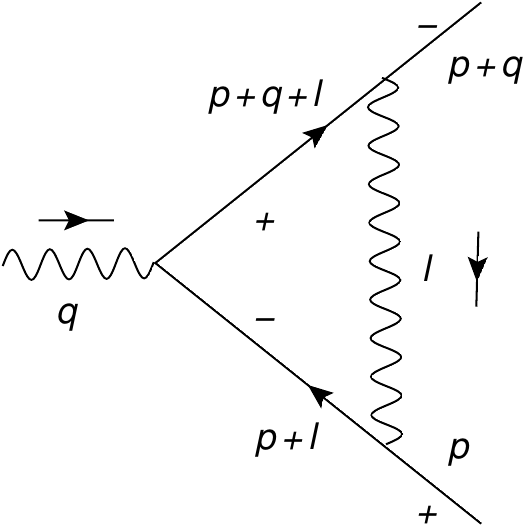}
\end{center}
\caption{The 1-loop contribution to the Boson-Fermion vertex.}
\label{bf}
\end{figure}
The expression for the diagram in Fig.~\ref{bf} can be written as,
\begin{equation}
\delta\Gamma^a_{\sigma\sigma'}(p,q)=(\tau^b\tau^a\tau^b)_{\sigma\sigma'}\lambda^3\int\frac{dl_\tau d^2{\bf l}}{(2\pi)^3}G_+^0(l+p+q)G_-^0(l+p)D(l).
\end{equation}
We now use the identity $\tau^b\tau^a=\delta^{ba}+i\epsilon^{bac}\tau^c$ twice to simplify the above expression. Then on defining $\delta\Gamma^a_{\sigma\sigma'}(p,q)=\tau^a_{\sigma\sigma'}\delta\Gamma(p,q)$, we have,
\begin{eqnarray}
\delta\Gamma(p,q)=-\frac{\lambda^3}{N}\int\frac{dl_\tau d^2{\bf l}}{(2\pi)^3}\bigg[\frac{1}{-i  (l_\tau+p_\tau+q_\tau)+(l_x+p_x+q_x)+(l_y+p_y+q_y)^2}\bigg]\nonumber\\ \bigg[\frac{1}{-i  (l_\tau+p_\tau)-(l_x+p_x)+(l_y+p_y)^2}\bigg]\bigg[\frac{1}{{\bf l}^2+r+[\Pi_0({\bf l})+\Pi_0(-{\bf l})+\Pi_0({\bf \tilde{l}})+\Pi_0(-{\bf \tilde{l}})]}\bigg].
\end{eqnarray}
Let us now evaluate this for zero external momenta $q=p=0$ at the critical point and check for singularities. The expression reduces to,
\begin{eqnarray}
\delta\Gamma=-\frac{\lambda^3}{N}\int\frac{dl_\tau d^2{\bf l}}{(2\pi)^3}\bigg[\frac{1}{-i   l_\tau+l_x+l_y^2}\bigg]\bigg[\frac{1}{-i   l_\tau-l_x+l_y^2}\bigg]\nonumber\\\bigg[\frac{1}{{\bf l}^2+\Pi_0({\bf l},l_\tau)+\Pi_0(-{\bf l},l_\tau)+\Pi_0({\bf \tilde{l}},l_\tau)+\Pi_0(-{\bf \tilde{l}},l_\tau)}\bigg].
\label{renver}
\end{eqnarray}
The above integrand has a very complex structure. Let us therefore analyze the (non-)singular nature of this diagram by power counting. One needs to be careful as the $\phi-$propagator has many combinations of powers of the internal momenta.
We begin by rescaling the variables as,
\begin{equation}
l_y=l, ~~l_x=l^2 l_x', ~~l_\tau=l^2l_\tau'
\end{equation}
Eqn. \ref{renver} then takes the form,
\begin{eqnarray}
\delta\Gamma&=&-\frac{\lambda^3}{N}\int\frac{dl_\tau'dl_x'dl}{(2\pi)^3}\bigg[\frac{1}{-i   l_\tau'+l_x'+1} \bigg]\bigg[\frac{1}{-i   l_\tau'-l_x'+1} \bigg] \nonumber \\
&&\times \frac{1}{l^2+[lf(l_x',l_\tau')+lf(-l_x',l_\tau')+\sqrt{l}g_+(l,l_x',l_\tau')+\sqrt{l}g_-(l,l_x',l_\tau')]/2\pi  },\\
f(l_x',l_\tau')&=&\sqrt{(1/2-l_x')+\sqrt{(1/2-l_x')^2+ l_\tau'^2}}, \\
g_\pm((l,l_x',l_\tau')&=&\sqrt{(l^3l_x'^2/2\pm1)+\sqrt{(l^3l_x'^2/2\pm1)^2+ l^2l_\tau'^2}}
\end{eqnarray}
All we need to do now is to check whether this expression (which is so far exact) is singular in the IR and UV. In the IR, we can ignore the momentum dependence of the fermionic Green's functions compared to $1$ in the denominator. Moreover, $f(l_x',l_\tau')\approx1$ and $g_+(l,l_x',l_\tau')\approx1, g_-(l,l_x',l_\tau')\approx0$ in this small momentum limit. Therefore, the above expression reduces to,
 \begin{eqnarray}
\delta\Gamma&=&-\frac{\lambda^3}{N}\int_0^\epsilon\frac{dl_\tau'dl_x'dl}{(2\pi)^3}\frac{1}{l^2+(2l+\sqrt{l})/2\pi  },
\end{eqnarray}
where $\epsilon$ is a small cutoff. But the above expression is convergent, so that there are no IR singularities.

Let us now check for UV singularities. We proceed by introducing a characteristic lower cutoff $\Lambda$ which is large, but finite, such that the integration runs from $\Lambda$ to $\infty$. Then, in the limit of these large momenta, we have,
\begin{eqnarray}
\delta\Gamma&=&-\frac{\lambda^3}{N}\int_\Lambda^\infty\frac{dl_\tau'dl_x'dl}{(2\pi)^3}\bigg[\frac{1}{-i   l_\tau'+l_x'} \bigg]\bigg[\frac{1}{-i   l_\tau'-l_x'} \bigg]\frac{1}{l^2+[l\sqrt{\sqrt{l_x'^2+  l_\tau'^2}+l_x'}+2l^2l_x']/2\pi  },\nonumber \\
\end{eqnarray}
 where we have ignored the contribution from $f(-l_x',l_\tau')$ compared to $f(l_x',l_\tau')$ and $  l^2l_\tau'^2$ compared to $l^6l_x'^4$. We first want to do the integral over $l_x'$ and $l_\tau'$. It is more convenient to change to polar coordinates, $(l_x',   l_\tau')\rightarrow(r,\theta)$. However, we estimate $l_x'\approx r$ and eliminate the $\theta$ dependence, which does not give rise to any singularities. Then,
\begin{equation}
\delta\Gamma=\frac{\lambda^3}{N}\int_\Lambda^\infty\frac{r dr dl}{(2\pi)^3}\frac{1}{r^2}\frac{1}{l^2+[l\sqrt{2r}+2l^2r]/2\pi  }.
\end{equation}
We also notice that $l^2, l^2r>l\sqrt{r}$ in the UV. Therefore, ignoring the $l\sqrt{r}$ term, we obtain,
 \begin{equation}
\delta\Gamma=\frac{\lambda^3}{N}\int_\Lambda^\infty\frac{dr dl}{(2\pi)^3}\frac{1}{r}\frac{1}{l^2+l^2r/\pi  }.
\end{equation}
Both the $r$ and $l$ integrals can be performed easily to verify that $\delta\Gamma$ is convergent in the UV. Therefore, the 1-loop vertex correction is non-singular both in the UV and in the IR.


\bibliographystyle{apsrev}

\end{document}